\documentclass[12pt,a4paper]{article}

\RequirePackage[l2tabu, orthodox]{nag}

\usepackage{mjheppub}

\usepackage[utf8]{inputenc}
\usepackage{amsmath,amsfonts,amssymb,epsfig,fp,youngtab,color,wasysym,bbm,hyperref}

\usepackage{tikz}
\usetikzlibrary{decorations.markings,arrows,decorations.pathreplacing,cd}
\usepackage{braids}

\usepackage{fullpage}

\def\be{\begin{equation}\begin{gathered}}
\def\ee{\end{gathered}\end{equation}}
\newcommand{\eq}[1]{\begin{equation}\begin{gathered}
#1\end{gathered}\end{equation}}
\newcommand{\eqs}[1]{\begin{equation}\begin{gathered}\begin{split}
#1\end{split}\end{gathered}\end{equation}}
\def\Sum{\sum\limits}
\def\Prod{\prod\limits}

\def\d{\partial}

\def\tr{\mathrm{\,tr\,}}
\def\bs{\boldsymbol}
\def\mf{\mathfrak}
\def\mc{\mathcal}

\def\Int{\int\limits}

\def\bpm{\begin{pmatrix}}
\def\epm{\end{pmatrix}}
\def\lb{\left(}
\def\rb{\right)}

\newcommand{\sign}{\operatorname{sign}}
\newcommand{\im}{\operatorname{Im}}

\numberwithin{equation}{section}
\newtheorem{theorem}{Theorem}

\newtheorem{proposition}{Proposition}

\def\monSpaceC{{\mc M_{0,4}^{SL(2,\mathbb C)}\left(\bs\theta\right)}}
\def\monSpaceR{{\mc M_{0,4}^{SU(2)}\left(\bs\theta\right)}}
\def\monSpace{{\mc M_{0,4}^{SL(2,\mathbb{C})}}}
\def\taut{\tau\!\tau}
\def\ttheta{{\bf\tilde{\bs\theta}}}

\title{Crossing invariant correlation functions at $c=1$ from isomonodromic $\tau$ functions}

\author{Pavlo  Gavrylenko  $^{1,2,3}$,}
\author{Raoul Santachiara   $^4$}

\affiliation{\vspace{2mm} $^1$ 
Center for Advanced Studies, Skolkovo Institute of Science and Technology,\\ 143026 Moscow, Russia
}

\affiliation{$^2$
 NRU HSE, International Laboratory of Representation Theory and Mathematical Physics,\\ 119048 Moscow, Russia}

\affiliation{$^3$ Bogolyubov Institute for Theoretical Physics,  03680 Kyiv, Ukraine}

\affiliation{$^4$ LPTMS, CNRS (UMR 8626), Univ Paris-Sud, Universit\'e Paris-Saclay, 91405 Orsay, 
France}

\emailAdd{
pasha.145@gmail.com, 
raoul.santachiara@gmail.com
}

\begin{document}

\abstract{We present an approach that gives rigorous construction of a class of crossing invariant functions in $c=1$ CFTs from the weakly invariant distributions on the moduli space  $\monSpace$ of $SL(2,\mathbb{C})$ flat connections on the sphere with four punctures.
By using  this approach we show how to obtain correlation functions in the Ashkin-Teller and the Runkel-Watts theory.
Among the possible crossing-invariant theories, we obtain also the analytic Liouville theory,  whose consistence was assumed only  on the basis of numerical tests.}

\maketitle

\section{Introduction} 
Bootstrap solutions of two dimensional CFTs are known since longtime  \cite{bpz84}.
 The existence and the form of these solutions depend strongly on the value of the central charge $c$, that parametrizes the current conformal algebra, the Virasoro algebra. 
When $c$ takes rational values smaller than one, $ c\in \mathbb{Q} \;\text{and}\; c<1$, one finds the {\it minimal  models} whose spectrum contains a finite number of degenerate Virasoro representations \cite{bpz84}.
 For $c$ irrational, $c\notin \mathbb{Q}$, we have the {\it generalized minimal models} with a diagonal (i.e. with only spinless primary fields) spectrum formed by a discrete but infinite number of degenerate representations \cite{bz06}. The correlation functions of the minimal and of the generalized minimal models are built from the solutions of Fuchsian differential equations and admit a (Coulomb gas) integral representation. This property makes the analytic proof of the crossing symmetry relatively simple. For central charge  $c \geq 25$, one finds the celebrated {\it Liouville field theory} that describes the two-dimensional quantum gravity \cite{zamlecture,Tecrev} and that has been recently related to the random energy models \cite{CRSL,CLRS18}. The Liouville field theory has a continuous and diagonal spectrum \cite{do94,zz95,te95} and its crossing symmetry has been proved in \cite{tes03b}. We notice also that the Liouville theory can be constructed by using a  (mathematically rigorous) probabilistic approach  \cite{vargasdozz17}, thus setting the consistency of this theory on very solid ground. The Liouville theory can be analytically continued to complex values of the central charge, $c\in \mathbb{C}-\{-\infty,1\}$. The general understanding  is that the analytic continuation of the Liouville theory is the unique bootstrap solution with continuous and diagonal spectrum in the region $c\in \mathbb{C}-\{-\infty,1\}$ \cite{RiSa15,CoKrLiYi17}. 

Despite the success of the above mentioned CFTs, there are many examples of critical points where a satisfying CFT description is lacking. We mention for instance the non-unitary critical systems with central charge $c \in \mathbb{R}_{\leq 1}$ such as the critical percolation \cite{psvd13,Saleur3point, PiRiSa16} or the random critical points \cite{DoPiPu95}. These models point out the existence of bootstrap solutions still to uncover.   
Motivated by this and by the success of  numerical bootstrap approaches in higher dimension CFT  \cite{epp+14}, the last few years have witnessed renewed efforts in the search of new CFTs. 
The outcome of this research unveiled indeed new solutions: some presenting  a discrete and non-diagonal spectrum \cite{EsIk15,PiRiSa16,MiRi17} and others with a continuous and diagonal spectrum \cite{RiSa15}.  
We refer to these latter solutions as the Liouville type solutions. In general, the crossing symmetry of these solutions is checked  only numerically,  mainly because the conformal blocks do not have integral representation nor satisfy differential equations.

The central charge $c=1$ represents a particularly interesting case where many different known CFTs exist. In this paper we consider in particular the Gaussian free field, the Ashkin-Teller model and two different Liouville type solutions: one is the {\it Runkel-Watt}s theory \cite{rw01} that can be found via a  $c\to 1$ limit  of the Liouville solution defined at $c\in \mathbb{C}-\{\infty,1\}$ \cite{sch03}. The other Liouville type solution is the one proposed in \cite{RiSa15} that is referred in this paper as the {\it analytic Liouville theory}. 

We will provide a derivation of the crossing symmetry of the $c=1$ theories by using the relation  between isomonodromic deformations and Virasoro  conformal blocks  \cite{gil12}. In this work we will fully exploit this  connection to provide a unifying theoretical framework to find a class of bootstrap solutions in $c=1$ theories. Among these solutions, we find  the analytic Liouville theory. This proves for the first time the crossing symmetry of this latter theory.

\section{Isomonodromic deformations -- Virasoro algebra relation}
We review here the connection between the theory of isomonodromic deformations of Fuchsian systems of differential equations and the $c=1$ Virasoro algebra conformal blocks.
\subsection{Rank 2 Fuchsian systems and the associated $\tau$ function}
 A rank $2$ Fuchsian system is defined by the system of the ordinary first order differential equations
\begin{equation}
\label{eq:FuchsSystem}
 \frac{dY(z)}{dz}=\left( \frac{A_0}{z}+\frac{A_{1}}{z-1}+\frac{A_t}{z-t}\right) Y(z)\;,
\end{equation}
where $Y(z)$ is $2\times 2$ matrix of fundamental solution and  $A_0,A_1,A_t$ are constant  $2\times 2$-matrices that, for simplicity, we consider semi-simple (i.e. diagonalizable). Moreover we can assume,  without loss of generality, the matrices $A_{\nu}$ to be traceless. Indeed, the addition operation, sending $
 (A_0,A_1,A_t)\to(A_0-\frac12\text{Tr} A_0 \cdot \mathbbm{1},A_1-\frac12\text{Tr} A_1 \cdot \mathbbm{1},A_t-\frac12\text{Tr} A_t \cdot \mathbbm{1})\;,
$ corresponds to the gauge transform $
 Y(z)\to Y(z)\cdot(z)^{-\frac12\text{Tr} A_0}(z-1)^{-\frac12\text{Tr} A_1}(z-t)^{-\frac12\text{Tr} A_t}$ for the system (\ref{eq:FuchsSystem}). We  can therefore assume that:
 
$$A_{\nu}\sim \text{diag}\left(\theta_{\nu}, -\theta_{\nu}\right),\quad \nu=0,1,t,$$
where $\sim$ is used for similar matrices. The residue matrix at $z=\infty$ is defined by 
$$
 A_{\infty}=-A_0-A_1-A_{t}\;\quad  A_{\infty}\sim \text{diag} \left(\theta_{\infty},-\theta_{\infty}\right).
$$
The monodromy group of the above system is characterized, in addition to the four parameters  $\bs\theta=\left(\theta_0,\theta_t,\theta_1,\theta_{\infty}\right)$, by the other two parameters that we denote as $\sigma$ and $s$. The monodromy data $(\bs \theta, \sigma, s)$ are considered as the coordinates on the monodromy manifold. We will usually  indicate a point of  this manifold by $P$,   $P= (\bs \theta, \sigma, s)$. A precise definition of the monodromy manifold  as well as an explicit parametrization of all its elements, is provided in section \ref{sec:moduli}. The rank 2 system (\ref{eq:FuchsSystem}) with four singularities at $z=0,1,t,\infty$ is not rigid \cite{katz1996rigid,BHS18}, i.e.  the monodromy data do not fix its form: there is one accessory parameter that one can deform by keeping the monodromy data invariant. Choosing the singularity position $t$ as the deformation parameter, isomonodromic deformations of the system \eqref{eq:FuchsSystem} are governed by the equation
\eq{
\frac{\d Y(z)}{\d t}=-\frac{A_t}{z-t}Y(z)\,.
}
Compatibility condition of the latter formula with \eqref{eq:FuchsSystem} gives the Schlesinger equations that describe the non-linear evolution of $\{A_\nu\}$ under the isomonodromic transformation:
\eq{
\frac{\d A_0}{\d t}=\frac{[A_t,A_0]}{t}\,,\qquad \frac{\d A_1}{\d t}=\frac{[A_t,A_1]}{t-1}\,,\qquad \frac{\d A_t}{\d t}=-\frac{[A_t,A_0]}{t}-
\frac{[A_t,A_1]}{t-1}\,.
}
There is a statement that there exists tau function defined by formula
\eq{
\frac{\d}{\d t}\log\tau(t)=\frac{\tr A_0A_t}{t}+\frac{\tr A_1A_t}{t-1}\,.
\label{eq:tau_definition}
}
The existence of a $\tau$ function is equivalent to the existence of a closed $1$-form on the space of the deformation parameters \cite{Jimbo81}. In the 4-singularities case, where there is one single deformation parameter, this is a trivial statement. However, for more than four singularities, the existence of a $\tau$ function becomes non-trivial.
Since isomonodromic deformations preserve monodromies, tau-function depends on $t$ and on monodromy data,
$\tau(t)=\tau\left(P;t\right)$.

\subsection{Virasoro $c=1$ conformal blocks}

We refer the reader to \cite{rib14} for the definition of the Virasoro algebra and its representation theory. 
An important special function associated to the Virasoro representations is the four-point conformal block.  Once we fix the central charge to one, $c=1$, this is a function of six parameters: the value of the cross-ratio $t$, the dimensions $\Delta_{\nu}$, $\nu=0,1,t,\infty$ of the four primary fields sitting at positions $z=0,1,t,\infty$, and the dimension $\Delta$ of the Virasoro representation that flows in the internal channel. 
We recall that the internal field is the field produced in the fusion of the primary  at $z=t$ with one of the other primaries sitting at $z=0,1,\infty$. In the $s-$channel ($t-$channel) conformal block $\mc B^{(s)}(\{\Delta_{\nu}\};\Delta;t)$ ($\mc B^{(t)}(\{\Delta_{\nu}\};\Delta;t)$), $\Delta$ is the dimension of the representation appearing in the fusion between the primaries with dimension $\Delta_t$ and $\Delta_0$ ($\Delta_{1}$): the corresponding conformal block is defined in a $t$ expansion ($1-t$ expansion). One has:
$$\mc B^{(t)}(\Delta_{0},\Delta_t,\Delta_1,\Delta_\infty;\Delta;t)=(1-t)^{\Delta_0+\Delta_\infty-\Delta_t-\Delta_1}\mc B^{(s)}(\Delta_1,\Delta_t,\Delta_0,\Delta_\infty;\Delta;1-t).$$ 
Henceforth, we will drop the index of channel $(s)$ or $(t)$ because we will always imply  the conformal blocks to be in the $s-$channel, $B^{(s)}(\{\Delta_{\nu}\};\Delta;t)\to B(\{\Delta_{\nu}\};\Delta;t)$.
To  make the link with the isomonodromic transformations, one  parametrizes the dimensions $\Delta$ of the primary fields with the momenta $\theta$ such that $\Delta=\theta^2$:
\eq{
\mc B(\Delta_0,\Delta_t,\Delta_1,\Delta_\infty;\Delta;t) \to \mc B(\bs \theta;\sigma;t),\\
\text{with}\;\Delta_{\nu} =\theta_{\nu}^2, \; (\nu=0,t,1,\infty), \quad \text{and} \; \Delta=\sigma^2,
}
From its definition, the   $\mc B(\bs \theta;\sigma;t)$ is manifestly invariant under the change of sign of  $\theta_{\nu}$ and of $\sigma$. 

\subsection{$\tau$ functions in terms of Virasoro $c=1$ conformal blocks}
The main result in \cite{gil12} is that the isomonodromic
tau-function (\ref{eq:tau_definition}) is a linear combination of conformal blocks with appropriate coefficients. In particular, the $\tau(P,t)$ function can be expressed as a series in $t$ via the function $\tau_{0t}(P,t)$: 
\eq{
\tau(P,t)=\tau_{0t}(\bs \theta,\sigma_{0t},s_{0t},t)=\Sum_{n\in\mathbb Z}s_{0t}^n
C(\bs\theta;\sigma_{0t}+n)\mc B(\bs\theta;\sigma_{0t}+n;t)\,,
\label{eq:isoVira}
}
where the constants $C(\bs\theta;\sigma_{0t})$ are expressed in terms of the Barnes $G-$functions:
\eq{
C(\bs\theta;\sigma_{0t})=\frac{\prod_{\epsilon,\epsilon'=\pm}G(1+\theta_t+\epsilon\theta_0+\epsilon'\sigma_{0t})
G(1+\theta_1+\epsilon\theta_\infty+\epsilon'\sigma_{0t})}{G(1+2\sigma_{0t})G(1-2\sigma_{0t})G(1+2 \theta_{0})G(1+2 \theta_{t})G(1+2 \theta_{1})G(1+2 \theta_{\infty})}\,.
\label{eq:C_GIL}
}
Differently from the conformal blocks, the constants $C(\bs\theta;\sigma_{0t})$ are not symmetric under the change of sign, $\theta_{\nu}\to -\theta_{\nu}$: this is reminiscent of the reflection coefficient in Liouville theory \cite{rib14,RVrc18}.
 
The form of the Fuchsian system (\ref{eq:FuchsSystem}) is preserved by global conformal ($SL(2,\mathbb C)$) map $z\mapsto (az+b)/(cz+d)$, and in particular by the map $z\mapsto 1-z$, which permutes $0$ with $1$ and maps $t$ to $1-t$. Using this map one finds another representation for the tau function as a series in $1-t$:
\eq{
\label{eq:tau1t}
\tau_{1t}(\tilde{P},t)=\Sum_{n\in\mathbb Z}s_{1t}^n
C(\tilde{\bs \theta};\sigma_{1t}+n)\mc B(\tilde{\bs\theta};\sigma_{1t}+n;1-t)\,.
}
In the above equation the point $\tilde{P}$:
\eq{
\label{eq:PtotildeP}
\tilde{P}=(\tilde{\bs\theta},\sigma_{1t},s_{1t}), \quad \tilde{\bs\theta}=(\theta_1,\theta_t,\theta_0,\theta_{\infty}) ,
} is the image of $P$ under the map $z\mapsto 1-z$, $P\to \tilde{P}$, see section \ref{sec:st_transformation}. 
Since \eqref{eq:tau_definition}
defines only logarithmic derivative of the tau function, it can get some non-trivial multiplicative factors after analytic continuation. This is actually the case for $\tau_{0t}(P,t)$ and $\tau_{1t}(\tilde{P},t)$ that are tau functions of the same system. The ratio
\eq{
\frac{\tau_{0t}(P,t)}{\tau_{1t}(\tilde{P},t)}=\chi_{01}
\label{eq:connection_ILTy}
}
is called (up to some
normalization) connection constant. Explicit formula for this constant has been conjectured in \cite{ilt13} and then rigorously proven in \cite{ILP16}.

In principle, functions $\tau_{0t}(P,t)$ for different values of monodromies contain all possible $c=1$ conformal blocks --- for example, they can
be extracted by the inverse Fourier transformation. This procedure was used in \cite{ilt13} to obtain the fusion kernel for arbitrary $c=1$ conformal blocks from connection constant \eqref{eq:connection_ILTy}. In principle, this formula for the fusion kernel could be useful for the study of crossing invariance in $c=1$ CFT. As a matter of fact, this turns out to be an extremely difficult approach because the fusion kernels are in general continous with a very subtle structure of singularities. 

Before entering into details, let us first give the general sketch of the approach we will use to solve this problem.  We advocate a different point of view: {\it the good basis in the space of $c=1$ conformal blocks are not the 
conformal blocks $\mc B(\bs\theta,\sigma_{0t},t)$ themselves, but isomonodromic tau functions $\tau_{0t}(P,t)$ or $\tau_{1t}(P,t)$}. The main reason for this choice is the much simpler formulas for the tau function crossing transformation \eqref{eq:connection_ILTy}. 
Another reason is based on the fact that the tau functions are common eigenvectors of the Verlinde loop operators that 
act on the space of conformal blocks and that commute in $c=1$ case \cite{ilt14}. The CFT correlation function will be expressed in the tau function basis and the associated bootstrap relations will be solved by exploiting the formulas for the connection constant  (actually even simpler consequence of it). We propose the four-point correlation function  $\mc F_{\bs \theta}(t,\bar t)$  to have this form:
$$
\mc F_{\bs \theta}(t,\bar t) \to \mc F(t,\bar t)[d \mu_{\bs \theta}(P)]\propto \int_{P} d\mu_{\bs \theta}(P)\tau_{0t}(P;t)\tau_{0t}(\iota(P),\bar t),
$$
where $\iota(P)$ is the image of $P$ under conjugation, $z\mapsto \bar{z}$, as explained in section \ref{sec:iota_tramsformation}. Therefore, given a distribution~\footnote{We call by distribution some measure without
positivity property and without normalization condition. One can also consider $d\mu_{\bs\theta}(P)$ as the element of the dual space to the (sub-)space of meromorphic functions of $P$, and the integral as pairing between function and distribution.}
 $d\mu_{\bs \theta}(P)$ on the monodromy data manifold, the correlation functions are averages of the product $\tau_{0t}(P;t)\tau_{0t}(\iota(P),\bar t)$ of holomorphic and anti-holomorphic tau functions. The analytic continuations in the cross ratio $t$ can be translated into the transformations of the monodromy manifold under the action of the braid group $\mathbb{B}_4$, which exchanges the positions of the primary fields. This is the case for instance for the transformation (\ref{eq:PtotildeP}), see section (\ref{sec:braid}),  which is associated to the continuation $t\to 1-t$. The invariance of the four-point correlation function depends therefore on the transformation properties of the  distribution $d\mu_{\bs \theta}(P)$ under  the braid group $\mathbb{B}_4$. As we will argue below, the  $c=1$ crossing symmetry solutions correspond to  distributions that are {\it weakly invariant}. The precise definition of weakly invariant distributions, as well as the weak equivalence between distributions,  are given  in (\ref{eq:weak_invariant}) and in (\ref{eq:weak_equivalence}).

In this paper we present several examples of weakly invariant distributions: one of them corresponds to  the  Runkel-Watts theory and another one to the  analytic Liouville theory \cite{RiSa15}, proving then conjecture about their crossing invariance for $c=1$. One
more example recovers particular correlation functions in Ashkin-Teller model, whose crossing symmetry was already proven in \cite{zam86at} by using the properties of elliptic functions.

\section{CFTs at $c=1$}
In this section we briefly review some CFT solutions at $c=1$.

\subsection{Gaussian free field}
The Gaussian free field (GFF) is described by a scalar field $\varphi(x)$ and the action $S[\varphi(x)]\sim\int d^2 x \; \left(\nabla \varphi(x)\right)^2$.  It is the simplest example of CFT as it is a free theory and the correlation functions can be computed via the Wick theorem. The primary fields are the exponential  fields $V_{\theta}(x)=e^{i \theta \varphi(x)}$ with $U(1)$ charge $\theta$ and conformal dimension $\Delta=\bar{\Delta}=\theta^2$.
One can demand unitarity and consider therefore the representations with positive dimension. The spectrum $\mathcal{S}^{GFF}$ of this CFT is:
\begin{align}
\label{eq:spect_RG}
\mathcal{S}^{GFF}: \quad &\{\Delta, \Delta\}, \quad \Delta= \theta^2 \nonumber\\ & \theta \in \mathbb{R}. 
 \end{align}
 For the four-point function, one obtains:
\eq{
\label{eq:corr_GFF}
\mathcal{F}^{GFF}_{\bs \theta}(t, \bar{t})=\left<V_{\theta_{0}}(0)V_{\theta_{t}}(t)V_{\theta_{1}}(1)V_{\theta_{\infty}}(\infty)\right> = \delta_{\theta_{0}+\theta_t+\theta_1+\theta_\infty,0}|t|^{4 \theta_{0}\theta_t}|1-t|^{4 \theta_{1}\theta_t},
} 
where the Kronecker symbol $\delta_{\theta_{0}+\theta_t+\theta_1+\theta_\infty,0}$ comes from the $U(1)$ charge symmetry. Using the fact that:
 $$\mc B(\theta_{0},\theta_t,\theta_1,-\theta_0-\theta_t-\theta_1;\theta_{0}+\theta_1;t)= t^{2\;\theta_{0}\theta_t}(1-t)^{2 \theta_1 \theta_t}, $$
the correlation (\ref{eq:corr_GFF}) can be factorized in the following way:
\eq{
\label{eq:corr_GFF_fact}
\mathcal{F}^{GFF}_{\bs \theta}(t, \bar{t})=\delta_{\theta_{0}+\theta_t+\theta_1+\theta_0,0}\; \mc B(\bs\theta;\theta_{0}+\theta_t;t)\mc B(\bs\theta;\theta_{0}+\theta_t;\bar{t}).
}

\subsection{The Ashkin-Teller model}
The Ashkin-Teller model \cite{SaleurAT} is a lattice spin model that can be defined as two coupled Ising models. The spin variables,  $s_1(\vec{j})$ and $s_2(\vec{j})$ are defined at each lattice site $\vec{j}=(j_x,j_y)$ and  can take each  two possible values $s_{1,2}(\vec{j})=0,1$. In addition to the  Ising couplings  $s_{1,2}(\vec{j}) s_{1,2}(\vec{j}+\vec{e}_{a})$, where $\vec{j}$ and $\vec{j}+\vec{e}_{a}$ are neighbouring sites,  there is a four spin (energy-energy) interaction  of type $s_{1}(\vec{j}) s_{1}(\vec{j}+\vec{e}_{a})s_{2}(\vec{j})s_{2}(\vec{j}+\vec{e}_{a})$. 
This model has a (self-dual) line of critical points that are described by a one parameter family of CFTs with central charge  $c=1$.  We denote this parameter with $N$. The study of the correlation function in this model has provided Virasoro $c=1$ solution \cite{zam86at} that we now briefly review. 

The (untwisted) spectrum contains the representations with dimension: 
\begin{align}
\mathcal{S}^{AT}:&\quad \{\Delta_{n,m},  \bar{\Delta}_{n,m}\}, \quad n,m \in \mathbb{Z}\nonumber \\ 
&\Delta_{n,m}=\left(\frac{n}{2N} +   m N\right ) ^2,\quad \bar{\Delta}_{n,m} = \left(\frac{n}{2N} - N \right) ^2,
\end{align}
This spectrum is therefore {\it discrete} and {\it non-diagonal}, with fields with scaling dimension $\lb\Delta+\bar\Delta\rb=n^2/(2N^2)+2m^2N^2$ and spin $\lb\Delta-\bar\Delta\rb=2mn$. The values of  $N=1/\sqrt{2}$ and $N=1$  correspond respectively to two decoupled Ising model, and to the four-states Potts model \cite{Wupotts}. The (twisted sector of the) spectrum contains also fields that have dimension $\Delta=1/16$ for any value of $N$. The set of parameters:
\eq{
\label{eq:Picard}
\bs\theta_{Picard}=\lb\frac14,\frac14,\frac14,\frac14\rb,
} 
corresponds to four  $\Delta=1/16$ representations. These representations are used in the computation of the magnetic correlations. There are four independent magnetic correlations \cite{DeVi11}, that can be classified according to their behavior  under operator position exchange: the invariant magnetic correlation $\langle s_{1} s_{1}s_{1}s_{1}\rangle=\langle s_{2} s_{2}s_{2}s_{2}\rangle$  and three magnetic correlations of type $\langle s_{1} s_{1} s_{2} s_{2}\rangle$ which transform one into another under this operation. The building blocks of such correlation are the functions $\mc B(\bs \theta_{Picard};\sigma_{0t},t)$ that have been computed exactly \cite{zam86at}:
\eq{
\label{eq:Picardblock}
\mc B\left(\bs\theta_{Picard};\sigma_{0t},t\right)= t^{-1/8}(1-t)^{-1/8}\frac{\left(16 q(t) \right)^{\sigma_{0t}^2}}{\vartheta_3(0|\eta(t))},
}
where $\eta(t)$ is a period of elliptic curve with branch-points $0,t,1,\infty$,  $q(t)=e^{i\pi\eta(t)}$ is  the elliptic nome.

The four different functions labeled by $(\epsilon,\epsilon')=(0,0), (0,1), (1,0), (1,1)$:
\begin{align}
\label{eq:ATcorr}
&\mc F^{AT,(\epsilon,\epsilon')}_{\bs \theta_{Picard}}(t,\bar{t})= \\&=\sum_{n,m\in \mathbb{Z}}(-1)^{\epsilon'm} (16)^{-\frac{(n+\frac\epsilon2)^2}{2N^2}-2m^2 N^2} \mc B\left(\bs\theta_{Picard};\frac{n+\frac\epsilon2}{2N} +   m N,t\right)\mc B\left(\bs\theta_{Picard};\frac{n+\frac\epsilon2}{2N} - Nm,\bar{t}\right) =\nonumber \\
&=\frac{|t|^{-1/4}|1-t|^{-1/4}}{|\vartheta_3(0|\eta(t))|^2} \sum_{n,m \in \mathbb{Z}} (-1)^{\epsilon'm}q(t)^{\Delta_{n+\frac\epsilon2,m}}\bar{q}(t)^{\bar{\Delta}_{n+\frac\epsilon2,m}},
\end{align}
were proposed to describe the invariant magnetic correlator $\langle s_{1} s_{1}s_{1}s_{1}\rangle$ (corresponding to $\epsilon=\epsilon'=0$),
and also three correlators that transform one to another under the crossing transformations: $\langle s_1s_2s_2s_1 \rangle$, $\langle s_1s_1s_2s_2 \rangle$, and $\langle s_1s_2s_1s_2 \rangle$, corresponding to $(\epsilon,\epsilon')=$ $(1,0)$, $(0,1)$, and $(1,1)$, respectively \cite{zam86at}.
The crossing symmetry of these functions can be proven by using the transformation properties of the elliptic functions. Notice that a generalization to any value of $c<1$ of the cases $(1,0)$, $(0,1)$, and $(1,1)$ has been proposed in \cite{PiRiSa16}.

\subsection{The Runkel-Watts theory}
\label{sec:RW}
In \cite{rw01}, a new $c=1$ solution was found by taking, via a subtle procedure, the limit $p\to \infty$ of the unitary Virasoro minimal series $\mathcal{M}_{p,p+1}$. The spectrum $\mathcal{S}^{RW}$ of this theory is {\it continuous} and {\it diagonal}: it contains all the non-degenerate fields with positive dimension $\Delta\geq 0$. It can be parametrized via  $\theta$ such that: 
\begin{align}
\label{eq:spect_RW}
\mathcal{S}^{RW}: \quad &\{\Delta, \Delta\}, \quad \Delta= \theta^2 \nonumber\\ & \theta \in \mathbb{R}-\frac{\mathbb{Z}}{2}. 
 \end{align}
The  three point structure constant $C^{RW}(\theta_1,\theta_2,\theta_3)$ factorizes into two terms:
\eq{
C^{RW}(\theta_1,\theta_2,\theta_3)=RW(\theta_1,\theta_2,\theta_3)\Phi(\theta_1,\theta_2,\theta_3)\,.
}
The term
\eq{
\label{eq:Phidef}
\Phi(\theta_1,\theta_2,\theta_3)=\frac{\Prod_{\epsilon_i=\pm}G(1+\epsilon_1\theta_1+\epsilon_2\theta_2+\epsilon_3\theta_3)}{
\Prod_{i=1}^3\Prod_{\epsilon=\pm}G(1+2\epsilon\theta_i)}
}
is analytic in $\theta_i$, $i=1,2,3$ while the term
\eq{
RW(\theta_1,\theta_2,\theta_3)=\left[\begin{array}{ll}0:\,
\frac{\sin\pi(\theta_2+\theta_3-\theta_1)\sin\pi(\theta_2+\theta_3+\theta_1)}{
\sin\pi(\theta_2-\theta_3-\theta_1)\sin\pi(\theta_2-\theta_3+\theta_1)}>0\,,\\
1:\, \frac{\sin\pi(\theta_2+\theta_3-\theta_1)\sin\pi(\theta_2+\theta_3+\theta_1)}{
\sin\pi(\theta_2-\theta_3-\theta_1)\sin\pi(\theta_2-\theta_3+\theta_1)}\leq 0\,.\end{array}\right.
\label{eq:RW}
}
is a step-function. 
The Runkel-Watts correlation function is:
\eq{
\mathcal{F}^{RW}_{\bs \theta}(t,\bar{t})= \int_{-\infty}^{\infty} d\, \sigma_{0t} \;C^{RW}(\theta_0,\theta_t,\sigma_{0t})C^{RW}(\sigma_{0t},\theta_1,\theta_{\infty}) B\left(\bs\theta;\sigma_{0t},t\right)\mc B\left(\bs\theta;\sigma_{0t},\bar t\right)\,.
\label{eq:RW_correlator}
}
Notice that the integration domain is over all $\sigma_{0t}\in \mathbb{R}$. This is the same as doing the integration over the spectrum (\ref{eq:spect_RW}) as the poles in
$\sigma_{0t}\in\frac12\mathbb Z$ are suppressed by the factor (\ref{eq:RW}). 

\subsection{The $c=1$  analytic  Liouville theory}

Motivated by the study of the geometrical properties in random Potts model, a Liouville type theory was proposed for $c\leq 1$ in \cite{RiSa15}.  The spectrum $\mathcal{S}^{AL}$ is continuous and diagonal. For $c=1$ one has 
\begin{align}
\label{eq:spect_AL}
\mathcal{S}_{c=1}^{AL}: \quad &\{\Delta, \Delta\}, \quad \Delta= \theta^2 \nonumber\\ & \theta \in \mathbb{R}. 
 \end{align} 
The structure constant $C^{c\leq 1}(\theta_1,\theta_2,\theta_3)$ is given in terms of product of Barnes double Gamma functions \cite{rib14}. At $c=1$ their expression simplify to
\eq{
C^{c=1}(\theta_1,\theta_2,\theta_3)=\Phi(\theta_1,\theta_2,\theta_3)\,.
}
Notice that, differently from the Runkel-Watts theory, there is no step function that suppresses the (non-integrable) divergences coming from the conformal blocks. The four point correlation function has therefore being defined by moving the contour in the complex plane:
\eq{
\label{eq:AL}
\mathcal{F}^{AL}_{\bs \theta}(t,\bar{t})= \int_{i h+ \mathbb{R}} d\, \sigma_{0t} \;\Phi(\theta_0,\theta_t,\sigma_{0t})\Phi(\sigma_{0t},\theta_1,\theta_{\infty}) \mc B\left(\bs\theta;\sigma_{0t},t\right)\mc B\left(\bs\theta;\sigma_{0t},\bar t\right),
}
where $h$ is an arbitrary constant. The property of the above integral will be discussed in detail below.
\section{Moduli space of flat connections}
\label{sec:moduli}

In this section we give self-contained description of the moduli space of $SL(2,\mathbb C)$ and $SU(2)$ flat connections on sphere with four
punctures. Almost all content of this section can be found as well in \cite{gil12}, \cite{ilt13}, \cite{ILP16} and \cite{GIL18}.

\subsection{Definitions}

Given some group $G$, the central object of our approach is the moduli space of $G$-flat connections~\footnote{Since we work only with
monodromies of the Fuchsian system, we will use only $G\subseteq GL(N,\mathbb C)$.} on sphere with 4 punctures: this is the space of all maps from the fundamental group of Riemann sphere with four punctures $\pi_1(\mathbb CP^1\setminus\{0,t,1,\infty\})$
to $G$, defined up to overall conjugation. Henceforth we will denote this space as $\mc M_{0,4}^G$, 
\eq{
\mc M_{0,4}^G={\rm Hom}\lb\pi_1(\mathbb CP^1\setminus\{0,t,1,\infty\}),G\rb/G\,.
}
The above map can be seen as a monodromy map that sends each path $\gamma \in \pi_1(\mathbb CP^1\setminus\{0,t,1,\infty\})$ to the monodromy of the solution of the Fuchsian system (equivalently the flat connection)  over this path \cite{yoshida1987fuchsian}.
The definition of $\mc M_{0,4}^G$ can be formulated in an equivalent and more explicit way in  terms of the set of equivalence classes of  monodromy matrices:
\eq{
\mc M_{0,4}^G=\{M_0,M_t,M_1,M_\infty|M_\nu\in G, M_0M_tM_1M_\infty=1\}/G\,.
} 
Two sets of matrices belong to the same equivalence class if they are related by an element $g$ of the group $G$, via an overall conjugation: $ \{M_\nu\}\mapsto\{g^{-1}M_\nu g\}$, $g\in G$. 

We consider here two particular spaces, one for $G=SL(2,\mathbb C)$ and another one for $G=SU(2)$. Moreover, it is convenient  to consider the submanifold 
$\mc M_{0,4}^G(\bs\theta) \subset \mc M_{0,4}^G$ by setting $M_{\nu}\sim \text{diag}(e^{2\pi i \theta_{\nu}},e^{-2\pi i \theta_{\nu}})$, i.e. by fixing the conjugacy classes $[\theta_{\nu}]$ --- 
\eq{
\monSpaceC=\\
=\{M_0,M_t,M_1,M_\infty|M_\nu\in SL(2,\mathbb C),\tr M_\nu=2\cos 2\pi\theta_\nu,M_0M_tM_1M_\infty=1\}/SL(2,\mathbb C)\,.
}
In this way we get symplectic leaves of the Poisson bracket \cite{goldman} on the moduli space of flat connections. Notice that the condition $\text{det} \;M_{\nu}=1$ follows from the tracelessness condition $\text{Tr} \;A_{\nu}=0$ for the residue matrices in (\ref{eq:FuchsSystem}).
In the $G=SU(2)$ the (special) unitarity condition $M_\nu\in SU(2)$ should also be imposed, so  we have 
$$\monSpaceR\subset\monSpaceC\subset \monSpace.$$ 

\subsection{Explicit construction of the $\monSpaceR$ and $\monSpaceC$ spaces}

We give here explicit coordinate description
of (the open charts of) $\monSpaceR$ and $\monSpaceC$. We suppose that $M_{0t}=M_0M_t$ is diagonalizable
(by unitary matrix in the unitary case):
\eq{
\label{eq:M0Mt}
M_0M_t=\bpm e^{2\pi i\sigma_{0t}}&0\\0&e^{-2\pi i\sigma_{0t}}\epm=e^{2\pi i\mf S_{0t}}\,.
}
The matrix $M_t$ and its inverse $M_t^{-1}$ can be represented as linear combinations of identity matrix and rank 1 matrix:
\eq{
\label{eq:Mt}
M_t=\bpm e^{-2\pi i\theta_{t}}&0\\0&e^{-2\pi i\theta_{t}}\epm +\lb e^{2\pi i\theta_t}-e^{-2\pi i\theta_t}\rb \bpm u_1  v_1& u_2v_1\\u_1 v_2 & u_2 v_2\epm\,,\\
M_t^{-1}=\bpm e^{2\pi i\theta_{t}}&0\\0&e^{2\pi i\theta_{t}}\epm +\lb e^{-2\pi i\theta_t}-e^{2\pi i\theta_t}\rb \bpm u_1  v_1& u_2 v_1\\u_1 v_2 & u_2v_2\epm
}
where $u_1, v_1,u_2, v_2 \in \mathbb{C}$ are complex numbers satisfying $u_1 v_1+u_2 v_2=1$. Using a more compact notation, $M_t=e^{-2\pi i\theta_{t}}\mathbbm{1}+\lb e^{2\pi i\theta_t}-e^{-2\pi i\theta_t}\rb u^{T}\otimes v $, $M_t^{-1}=e^{2\pi i\theta_{t}}\mathbbm{1}+\lb e^{-2\pi i\theta_t}-e^{2\pi i\theta_t}\rb u^{T}\otimes v $, where $u$ and $v$ are row vectors, $u=(u_1,u_2)$ and $v=(v_1,v_2)$. We further constraint the four variables $u_1,u_2,v_1,v_2$ by imposing the condition  $M_0=\text{diag}(e^{2i\pi \theta_0},e^{-2i\pi \theta_0})$.
Using the formula $M_0=M_{0t}M_t^{-1}$ and $\det(\mathbbm{1}+AB)=\det(\mathbbm{1}+BA)$ we get:
\eq{
\det\lb\lambda-M_0\rb=\lb \lambda-e^{2\pi i\theta_0}\rb\lb\lambda-e^{-2\pi i\theta_0}\rb=\\=\det\lb\lambda-e^{2\pi i(\mf S_{0t}+\theta_t)}+
\lb e^{2\pi i\theta_t}-e^{-2\pi i\theta_t}\rb e^{2\pi i\mf S_{0t}}u^T\otimes v\rb=\\=
\lb \lambda-e^{2\pi i(\theta_t+\sigma_{0t})}\rb\lb\lambda-e^{2\pi i(\theta_t-\sigma_{0t})}\rb\lb 1+
\lb e^{2\pi i\theta_t}-e^{-2\pi i\theta_t}\rb v\frac{e^{2\pi i\mf S_{0t}}}{\lambda-e^{2\pi i(\mf S_{0t}+\theta_t)}}u^T \rb
}
Computing this expression near the fake singularities of the right side we get two relations:
\eq{
u_1v_1=e^{-2\pi i\sigma_{0t}}\frac{\lb e^{2\pi i(\theta_t+\sigma_{0t})}-e^{2\pi i\theta_0}\rb\lb e^{2\pi i(\theta_t+\sigma_{0t})}-e^{-2\pi i\theta_0}\rb}{
\lb e^{2\pi i\theta_t}-e^{-2\pi i\theta_t}\rb\lb
e^{2\pi i(\theta_t+\sigma_{0t})}-e^{2\pi i(\theta_t-\sigma_{0t})}\rb}=\\=
\frac{\sin\pi(\theta_t+\sigma_{0t}-\theta_0)
\sin\pi(\theta_t+\sigma_{0t}+\theta_0)}{\sin 2\pi\theta_t \sin 2\pi\sigma_{0t}}\,,\\
u_2v_2=e^{2\pi i\sigma_{0t}}\frac{\lb e^{2\pi i(\theta_t-\sigma_{0t})}-e^{2\pi i\theta_0}\rb\lb e^{2\pi i(\theta_t-\sigma_{0t})}-e^{-2\pi i\theta_0}\rb}{
\lb e^{2\pi i\theta_t}-e^{-2\pi i\theta_t}\rb\lb
e^{2\pi i(\theta_t-\sigma_{0t})}-e^{2\pi i(\theta_t+\sigma_{0t})}\rb}=\\=-
\frac{\sin\pi(\theta_t-\sigma_{0t}-\theta_0)
\sin\pi(\theta_t-\sigma_{0t}+\theta_0)}{\sin 2\pi\theta_t \sin 2\pi\sigma_{0t}}\,.
}
The four variables $u_1, u_2, v_1, v_2$ are defined modulo the transformation
$u,v \mapsto \lambda u, \lambda^{-1}v$. We can thus set one of these coefficients to one, say $v_2=1$. The remaining three coefficients satisfy two conditions: it suffices then 
to parametrize these four variables by one parameter  $s_{0t}^A$:
\eq{
u_1=\frac{\sin\pi(\theta_t+\sigma_{0t}-\theta_0)\sin\pi(\theta_t+\sigma_{0t}+\theta_0)}{\sin2\theta_t\sin2\pi\sigma_{0t}}e^{-2\pi i \sigma_{0t}}(s^A_{0t})^{-1}\,,\quad
v_1=s_{0t}^A e^{2\pi i\sigma_{0t}}\,,\\
u_2=-\frac{\sin\pi(\theta_t-\sigma_{0t}-\theta_0)\sin\pi(\theta_t-\sigma_{0t}+\theta_0)}{\sin2\pi\theta_t\sin2\pi\sigma_{0t}}\,,\quad
v_2=1\,.
}
Now we check the unitarity condition for matrix $M_t$: $M_t^{-1}=M_t^\dagger$. This leads to the condition $u_iv_j=u_j^*v_i^*$~\footnote{
which actually implies that $u=\alpha v^*$ for some $\alpha\in\mathbb R_+$}, which gives nontrivial relation for $i=1, j=2$:
\eq{
|s_{0t}^A|^2=-\frac{\sin\pi(\theta_t+\sigma_{0t}-\theta_0)\sin\pi(\theta_t+\sigma_{0t}+\theta_0)}{\sin\pi(\theta_t-\sigma_{0t}-\theta_0)\sin\pi(\theta_t-\sigma_{0t}+\theta_0)}
\label{eq:unitarity1}
}
The above relation can give solutions for $s_{0t}^A$ only when the r.h.s. is positive, which imposes non-trivial constraint on $\sigma_{0t}$, like spherical triangle
inequality: the step function $RW(\theta_t,\sigma_{0t},\theta_0)$,  defined in (\ref{eq:RW}), should equal to one. Moreover, obviously all $\theta$- and
$\sigma$-variables
should be real: $\theta_\nu\in\mathbb R$, $\sigma_{\mu\nu}\in\mathbb R$.
The same conditions should hold for the other triple of matrices, so the manifold
$\monSpaceR$ will be defined by reality constraints and two conditions:
\eq{
RW(\theta_0,\theta_t,\sigma_{0t})=1\quad \text{ and }\quad RW(\theta_1,\theta_\infty,-\sigma_{0t})=1\,.
\label{eq:unitarity}
}
The above conditions fix the fusion rules of the Runkel-Watts theory. In \cite{rw01} these fusion rules have been derived from a subtle limit of the minimal model fusion rules. It is quite remarkable that the same conditions can be associated to the monodromy submanifold $\monSpaceR$.

So far we have constructed monodromy matrices ($M_0$, $M_t$) only for one half of the sphere with 4 punctures, which is sphere with 3 punctures. We see
from this construction that $\mc M_{0,3}^{SL(2,\mathbb C)}(\bs\theta)$, as well as $\mc M_{0,3}^{SU(2)}(\bs\theta)$, are just points: all monodromy matrices
are defined only by the eigenvalues of $M_0, M_t, M_{0t}$, and $s_{0t}^i$ can be removed by the overall diagonal conjugation (in the $SU(2)$ case its modulus
is fixed, but phase can be removed). In other words, we see well-known fact that the second order Fuchsian system with three singulariries is {\it rigid}, i.e. the local exponents $\theta_0,\theta_t$ and $\sigma_{0t}$ are enough to parametrize the monodromy manifold.  

Coming back to our problem, we can construct the remaining monodromies, $M_{1},M_{\infty}$  using the same method and by  replacing  $\theta_0\to\theta_1$,
$\theta_t\to\theta_\infty$, $\sigma_{0t}\to -\sigma_{0t}$ (remember that $M_{1\infty}=M_{0t}^{-1}$), $s_{0t}^A\to s_{0t}^B e^{2\pi i\sigma_{0t}}\frac{\sin\pi(\theta_\infty-\theta_1-\sigma_{0t})}{
\sin\pi(\theta_\infty-\theta_1+\sigma_{0t})}$,  exactly as it was done in \cite{ILP16}:
\eq{
M_\infty=e^{-2\pi i\theta_\infty}\mathbbm 1+\lb e^{2\pi i\theta_\infty}-e^{-2\pi i\theta_\infty}\rb \tilde u^T\otimes\tilde v,\quad
M_1=e^{-2\pi i \mf S}M_\infty^{-1}\,,
}
where
\eq{
\tilde u_1=-\frac{\sin\pi(\theta_\infty+\sigma_{0t}-\theta_1)\sin\pi(\theta_\infty-\sigma_{0t}+\theta_1)}{\sin2\theta_\infty\sin2\pi\sigma_{0t}}(s^B_{0t})^{-1}\,,\quad
\tilde v_1=s_{0t}^B \frac{\sin\pi(\theta_\infty-\theta_1-\sigma_{0t})}{
\sin\pi(\theta_\infty-\theta_1+\sigma_{0t})}\,,\\
\tilde u_2=\frac{\sin\pi(\theta_\infty+\sigma_{0t}-\theta_1)\sin\pi(\theta_\infty+\sigma_{0t}+\theta_1)}{\sin2\pi\theta_\infty\sin2\pi\sigma_{0t}}\,,\quad
\tilde v_2=1\,.
}
 Computing matrix elements explicitly we recover (up to simultaneous transposition) formulas
(3.32a-d) from \cite{ILP16}. All matrices are defined up to the overall conjugation, so the point on monodromy manifold depends only on
\eq{
s_{0t}=\frac{s_{0t}^A}{s_{0t}^B}\,.
\label{eq:s_ratio}
}
In this way we constructed explicit uniformization of $\monSpaceC$ by the two coordinates, $\sigma_{0t}$ and $s_{0t}$. It is important to observe that
$\monSpaceC$ is a holomorphic symplectic manifold with holomorphic symplectic form $d\omega$ obtained from Goldman bracket \cite{goldman} (or from Atiyah-Bott form), and $\sigma_{0t}, \log s_{0t}$ are its Darboux coordinates \cite{NRS11}:
\eq{
d\omega=\frac1{2\pi i} d\sigma_{0t}\wedge\frac{d s_{0t}}{s_{0t}}
\label{eq:AtiyahBott}
}

From the \eqref{eq:s_ratio} and \eqref{eq:unitarity1}, in the unitary case we have:
\eq{
|s_{0t}|^2=R^{unitary}(\sigma_{0t},\bs\theta)=\\
=\frac{\sin\pi(\theta_t+\sigma_{0t}-\theta_0)\sin\pi(\theta_t+\sigma_{0t}+\theta_0)}{\sin\pi(\theta_t-\sigma_{0t}-\theta_0)\sin\pi(\theta_t-\sigma_{0t}+\theta_0)}
\frac{\sin\pi(\theta_1+\sigma_{0t}-\theta_\infty)\sin\pi(\theta_1+\sigma_{0t}+\theta_\infty)}{\sin\pi(\theta_1-\sigma_{0t}-\theta_\infty)\sin\pi(\theta_1-\sigma_{0t}+\theta_\infty)}\,.
\label{eq:Runitary}
}
\subsection{Algebro-geometric description}

We have seen that the manifold $\monSpace$ is a six-dimensional space. This can be seen also by considering the space of invariant functions on it, that are given by traces of all possible products of matrices. Such  products are associated to  elements (paths) of
$\pi_1(\mathbb CP^1\setminus\{0,t,1,\infty\})$. Working in this geometric representation we can use the relation in $SL(2,\mathbb C)$
\eq{
A+A^{-1}=\mathbbm 1\cdot \tr A
}
to reduce any path to a combination of non self-intersecting paths. It is possible because this relation is actually a ``skein relation'' shown
in Figure~\ref{fig:skein}, so one application of it reduces the number of self-intersections by 1.
\begin{figure}[h!]
\begin{center}
\begin{tikzpicture}[scale=0.5]
\draw(-1,0) to[out=45,in=0] (0,3) to[out=180,in=135] (1,0);
\draw[xshift=4cm](-1,0) to[out=45,in=180] (0,3) to[out=0,in=135] (1,0);
\draw[xshift=8cm](-1,0) to[out=45,in=180] (0,0.5) to[out=0,in=135] (1,0);
\draw[xshift=8cm] (0,2.3) circle[radius=0.7];

\node at(2,1.5){$=\,\, -$};
\node at(6,1.5){$+$};
\node at(0.3,3.5){$A^{-1}$};
\node at(4,3.5){$A$};
\node at(8.5,3.5){$\mathbbm{1}\cdot\tr A$};
\end{tikzpicture}
\end{center}
\caption{Geometric interpretation of $A+A^{-1}=\mathbbm 1\cdot \tr A$.}
\label{fig:skein}
\end{figure}

As a basis of non self-intersecting cycles first we should take cycles around one point --- they give four functions $p_\mu=\tr M_\nu$. And second
we should take cycles around pairs of points: there are three such inequivalent cycles encircling $0t$, $1t$ and $01$, and they give rise to
three functions  $p_{\mu\nu}=\tr M_\mu M_\nu$ for $\mu\nu = 0t, 1t, 01$, which are nevertheless not independent. There is one non-trivial relation called 
Jimbo-Fricke affine cubic $W=0$:
\eq{
\label{eq:Wcubic}
W=p_{0t}p_{1t}p_{01}+p_{0t}^2+p_{1t}^2+p_{01}^2-p_{0t}\lb p_0p_t+p_1p_\infty\rb-p_{1t}\lb p_1p_t+p_0p_\infty\rb-\\-p_{01}\lb p_0p_1+p_tp_\infty\rb+
p_0p_tp_1p_\infty+p_0^2+p_t^2+p_1^2+p_\infty^2-4
}

So algebro-geometric description of our monodromy manifold looks as follows: 
\eq{
\monSpace={\rm Spec}\lb\mathbb C[p_{0t},p_{1t},p_{01},p_0,p_t,p_1,p_\infty]/(W)\rb
}

Let us express  the quantities $p_{\mu \nu}$ in terms of the variables $(\sigma_{0t},s_{0t})$ introduced in (\ref{eq:M0Mt}) and in (\ref{eq:s_ratio}). We have:
\eq{
p_{0t}=\tr M_0M_t=2\cos2\pi\sigma_{0t}\,,
\\
p_{1t}=\tr M_1M_t=\frac1{\sin^2 2\pi\sigma_{0t}}\lb\frac 12\lb p_tp_1+p_0p_\infty\rb-\frac14\lb p_0p_1+p_tp_\infty\rb p_{0t}-\right.\\
\left.- 4 \Sum_{\epsilon'=\pm}s_{0t}^{\epsilon'}\Prod_{\epsilon=\pm}\sin \pi(\epsilon\theta_0+\theta_t-\epsilon'\sigma_{0t})\sin\pi(\epsilon\theta_\infty+\theta_1-\epsilon'\sigma_{0t})\rb\,,
\\
p_{01}=\tr M_0M_1=\frac1{\sin^2 2\pi\sigma_{0t}}\lb\frac12\lb p_0p_1+p_tp_\infty\rb-\frac14\lb p_tp_1+p_0p_\infty\rb p_{0t}+ \right.\\
\left.+ 4 \Sum_{\epsilon'=\pm}s_{0t}^{-\epsilon'} e^{2\pi i\sigma_{0t}\epsilon'}\Prod_{\epsilon=\pm} \sin \pi(\epsilon\theta_0+\theta_t+\epsilon'\sigma_{0t})\sin\pi(\epsilon\theta_\infty+\theta_1+\epsilon'\sigma_{0t})\rb\,.
\label{eq:p0t}
}
One can verify that (\ref{eq:p0t}) and (\ref{eq:p1t}) are invariant under
$\theta_\nu\mapsto\theta_\nu+k_\nu$, $\sigma_{\mu \nu}\mapsto\sigma_{\mu \nu}+k_{\mu\nu}$,
$k_\nu, k_{\mu\nu}\in\mathbb Z$. These are quite obvious symmetries since the monodromy properties of a Fuchs solution depend on the local exponents (i.e. the eigenvalues of the residue matrices $A_{\nu}$) modulo an integer. There are also simple symmetries $\theta_\nu\mapsto-\theta_\nu$ for $\nu=0,\infty$. Their
analogs for $\nu=t,1$ are less trivial. For example, $\theta_t\mapsto-\theta_t$, $s_{0t}\mapsto s_{0t}\prod\limits_{\epsilon=\pm}\frac{\sin\pi(\epsilon\theta_0-\theta_t+\sigma_{0t})}{\sin\pi(\epsilon\theta_0+\theta_t+\sigma_{0t})}$. Combining all such symmetries together we may construct transformation $\varkappa$:
\eq{
\varkappa\left(\bs\theta,\sigma_{0t},s_{0t}\right)=\left(-\bs\theta,\sigma_{0t},s_{0t}
\cdot\prod_{\epsilon=\pm}\frac{\sin\pi(\theta_1-\sigma_{0t}+\epsilon\theta_\infty)
\sin\pi(\theta_t-\sigma_{0t}+\epsilon\theta_0)}{
\sin\pi(\theta_1+\sigma_{0t}+\epsilon\theta_\infty)\sin\pi(\theta_t+\sigma_{0t}+\epsilon\theta_0)}\right)\,,
\label{eq:kappa}
}
which will be used in the next subsection.
Other symmetries will be discussed below, in section (\ref{sec:st_transformation}).

\subsection{Involutions $\iota$ and $\iota'$}
\label{sec:iota_tramsformation}

The crossing symmetry of  correlation function depends on the transformation properties of the holomorphic and anti-holomorphic conformal blocks. Consider an analytic  function $\phi(z,P)$ whose monodromies are described by the point $P\in\monSpace$. It is therefore  natural to ask what are the monodromies of anti-analytic function $\phi(\bar z,P)$. 

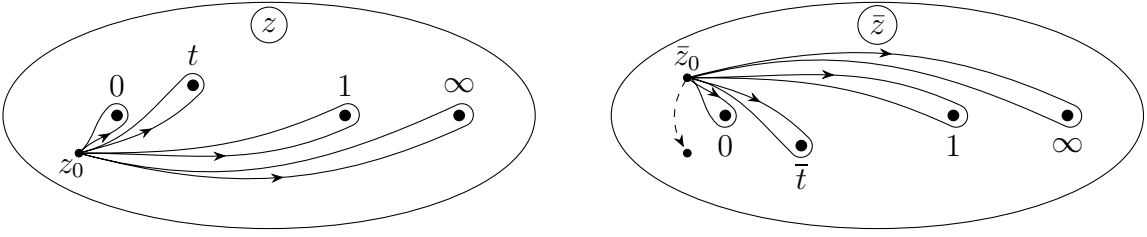
\begin{figure}[h!]
\begin{center}

\begin{tikzpicture}[decoration={markings,mark=at position 0.25 with {\arrow{Stealth}}}]
\draw(0,0) ellipse [x radius=3.5cm,y radius = 1.5cm];
\draw[fill=black,radius=0.07](-2,0) circle [radius=0.07]
(-1,0.4) circle[radius=0.07]
(1,0) circle[radius=0.07]
(2.5,0) circle[radius=0.07];

\node[inner sep=0] (z0) at (-2.5,-0.5){}; 
\node at (-2.6,-0.7){$z_0$};

\draw[postaction={decorate}] (z0) to[out=45,in=-135] (-1.9,-0.1) arc [radius=0.15cm,start angle=-45,end angle=135] to[out=-135,in=45] (z0);

\draw[postaction={decorate}] (z0) to[out=15,in=-135] (-0.9,0.3) arc [radius=0.15cm,start angle=-45,end angle=135] to[out=-135,in=15] (z0);

\draw[postaction={decorate}] (z0) to[out=0,in=-155] (1.1,-0.13) arc [radius=0.15cm,start angle=-65,end angle=115] to[out=-155,in=0] (z0);

\draw[postaction={decorate}] (z0) to[out=-15,in=-155] (2.6,-0.13) arc [radius=0.15cm,start angle=-65,end angle=115] to[out=-155,in=-15] (z0);

\draw[fill=black] (z0) circle[radius=0.05];

\node at (-2,0.4) {$0$};
\node at (-1,0.8) {$t$};
\node at (1,0.4) {$1$};
\node at (2.5,0.4) {$\infty$};

\node[draw,circle,inner sep=2] at (0,1.2){$z$};

\begin{scope}[xshift=8cm, yscale=-1]

\node[inner sep=0] (z0) at (-2.5,-0.5){}; 
\node[inner sep=2] (z1) at (-2.5,0.5){}; 
\node at (-2.5,-0.75){$\bar z_0$};

\draw[postaction={decorate}] (z0) to[out=45,in=-135] (-1.9,-0.1) arc [radius=0.15cm,start angle=-45,end angle=135] to[out=-135,in=45] (z0);

\draw[postaction={decorate}] (z0) to[out=15,in=-135] (-0.9,0.3) arc [radius=0.15cm,start angle=-45,end angle=135] to[out=-135,in=15] (z0);

\draw[postaction={decorate}] (z0) to[out=0,in=-155] (1.1,-0.13) arc [radius=0.15cm,start angle=-65,end angle=115] to[out=-155,in=0] (z0);

\draw[postaction={decorate}] (z0) to[out=-15,in=-155] (2.6,-0.13) arc [radius=0.15cm,start angle=-65,end angle=115] to[out=-155,in=-15] (z0);

\draw(0,0) ellipse [x radius=3.5cm,y radius = 1.5cm];
\draw[fill=black,radius=0.07](-2,0) circle [radius=0.07]
(-1,0.4) circle[radius=0.07]
(1,0) circle[radius=0.07]
(2.5,0) circle[radius=0.07];

\draw[fill=black] (z0) circle[radius=0.05];

\node at (-2,0.4) {$0$};
\node at (-1,0.8) {$\overline{t}$};
\node at (1,0.4) {$1$};
\node at (2.5,0.4) {$\infty$};

\node[draw,circle,inner sep=2] at (0,-1.2){$\bar z$};

\draw[-Stealth,dashed] (z0) to[bend left] (z1);

\draw[fill=black] (z1) circle[radius=0.05];

\end{scope}

\end{tikzpicture}

\end{center}
\caption{Complex conjugation of loops.}
\label{fig:conjugation}
\end{figure}

The involution of the loops under the map $z\mapsto \bar z$ is denoted by $\iota$, $\gamma\to\gamma^{\iota}$. The involutions of the generating elements of $\pi_{1}(\mathbb{C}P^1\ \{0,t,1,\infty\})$ are shown on
Figure~\ref{fig:conjugation}. One can see that the base point is shifted, so to compare the monodromies one has to move it to initial position
over some path~\footnote{
Different paths will give monodromies that differ by overall conjugation, so correspond to the same point of the monodromy manifold.}
drawn with dashed line in Figure~\ref{fig:conjugation}.
Moreover, consider for instance the loop $\gamma_t$ that is encircling  the point $t$ and  passes {\it under} the point $0$. The transformed loop $\gamma^{\iota}_t$ is inverted and  passes {\it over} the same point $0$. Therefore, taking into account the branch cut associated to the singularity in $0$, the loops $\gamma^{\iota}_t$ and $\gamma_t$ live on different sheets. The loop $\gamma^{\iota}_t$ is topologically equivalent to the loop product $\gamma_{0}^{-1}\gamma_{t}^{-1}\gamma_{0}$.

 Let us consider the action of the involution $\iota$ on the  matrices in $\monSpace$. This action (defined up to conjugation)  will be denoted by $\iota(M_\nu)$ or, equivalently, by  $M_\nu^\iota$. On the basis of the above considerations, one has:
\eq{
M_0^\iota=M_0^{-1}\,,\quad
M_t^\iota=M_0M_t^{-1}M_0^{-1}\,,\quad
M_1^\iota=M_0M_tM_1^{-1}M_t^{-1}M_0^{-1}\,,\\
M_\infty^\iota=M_0M_tM_1M_\infty^{-1}M_1^{-1}M_t^{-1}M_0^{-1}\,.
}

Now using the following identity in $SL(2,\mathbb C)$~\footnote{
We use the conjugation by matrix with $\det\neq1$, but its determinant can be easily fixed by introducing normalization factor, which obviously cancels.
}
\eq{
\bpm a&b\\c&d\epm^{-1}=\bpm 0&b\\-c&0\epm^{-1} \bpm a&b\\c&d\epm \bpm 0&b\\-c&0\epm\label{eq:SL_2_inverse}
}
and the fact that the matrices $M^{\iota}_{\nu}$ are defined up to overall conjugation, we can introduce new involution $\iota'$, whose
action is given by:
\eq{
M_0^{\iota'}=N_0M_0^{-1}N_0^{-1}\,,\quad
M_t^{\iota'}=N_0M_0M_t^{-1}M_0^{-1}N_0^{-1}\,,\quad
M_1^{\iota'}=N_0M_0M_tM_1^{-1}M_t^{-1}M_0^{-1}N_0^{-1}\,,
}
where we introduce
\eq{
N_0=\bpm0&(M_0)_{12}\\-(M_0)_{21}&0\epm,\quad N_1=\bpm0&(M_1)_{12}\\-(M_1)_{21}&0\epm\,.
}
Using \eqref{eq:SL_2_inverse} we get $M_0^\iota=M_0$ by definition, $M_0^\iota M_t^\iota=N_0M_{0t}^{-1}N_0^{-1}=M_{0t}$ by simple computation. Therefore
$M_t^{\iota'}=M_t$, as well. It remains to compute the action on $M_1$:
\eq{
M_1^{\iota'}=(N_0M_{0t}N_1^{-1}) M_1 (N_0M_{0t}N_1^{-1})^{-1}\,.
}
Matrix $N_0M_{0t}N_1^{-1}$ is diagonal, so diagonal elements of $M_1$ remain unchanged. Compute now one non-diagonal element:
\eq{
\iota'(M_1)_{12}=(M_1)_{12}\frac{(M_1)_{21}}{(M_1)_{12}}\frac{(M_{0t})_{22}}{(M_{0t})_{11}}\frac{(M_0)_{12}}{(M_0)_{21}}\,.
}
In other words it means that
\eq{
\iota'(\bar u_1 \bar v_2)=\bar u_1\bar v_2\cdot e^{4\pi i\sigma_{0t}}\frac{\bar u_2\bar v_1}{\bar u_1\bar v_2}\frac{u_1 v_2}{u_2 v_1}\,,
}
or equivalently
\eq{
\iota'(s_{0t})=\frac1{s_{0t}}\prod_{\epsilon=\pm}\frac{\sin\pi(\theta_1+\sigma_{0t}+\epsilon\theta_\infty)
\sin\pi(\theta_t+\sigma_{0t}+\epsilon\theta_0)}{
\sin\pi(\theta_1-\sigma_{0t}+\epsilon\theta_\infty)\sin\pi(\theta_t-\sigma_{0t}+\epsilon\theta_0)}\,.
\label{eq:involution}
}
Resuming, under an involution $\iota$, a point $P=\left(\bs \theta,\sigma_{0t}, s_{0,t}\right)$, $P\in \monSpace$ is transformed like:
\eq{
\label{eq:iotaP1}
P= \left(\bs \theta,\sigma_{0t}, s_{0t}\right)\mapsto \iota'(P)= \left(\bs \theta,\sigma_{0t},\iota'(s_{0t})\right)
}
We may also check that $\iota'(p_\nu)=\iota(p_\nu)=p_\nu$, as well as $\iota'(p_{0t})=\iota(p_{0t})=p_{0t}$ and $\iota'(p_{1t})=\iota(p_{1t})=p_{1t}$,
but $p_{01}$ changes:
\eq{
\label{eq:iotap01}
\iota'(p_{01})=\iota(p_{01})=\tr M_0^{-1}M_tM_1^{-1}M_t=p_tp_\infty+p_0p_1-p_{0t}p_{1t}-p_{01}\,.
}
This is noting but permutation of two sheets of Jimbo-Fricke surface \eqref{eq:Wcubic}. The same results have been obtained in \cite[p. 14]{ilt13}. Up to small misprint~\footnote{in \cite{ilt13} one has to replace $\omega_{0t}$ with $\omega_{01}$ in this formula}, the reader can compare the above formulas with the ones found in \cite{ilt13} (notice that in \cite{ilt13} the primed  quantities correspond to the ones transformed under involution. For instance,  $\iota(p_{01})$ here  corresponds to $p_{01}'$ in \cite{ilt13}).
It is also interesting to notice that involution $\iota$ coincides with the $y$-generator of extended modular group from the paper \cite{LTy08},
and to get the whole modular group one has to add two very similar generators.

Here during the construction of $\iota'$ we have used the fact that in $SL(2,\mathbb C)$ matrix and its inverse lie in the same conjugacy class 
\eqref{eq:SL_2_inverse}, so we could keep $\bs\theta$ the same.
In the general case one should have $\bs\theta\mapsto-\bs\theta$, so it is much
more natural to combine $\iota'$ with the transformation $\varkappa$ from \eqref{eq:kappa}: $\iota=\varkappa\circ\iota'$. Its action is much simpler:
\eq{
\iota(\bs\theta,\sigma_{0t},s_{0t})=(-\bs\theta,\sigma_{0t},s_{0t}^{-1})\,.
\label{eq:iotaP}
}
We remind that both transformations have the same action on the invariant functions, the only difference is the action on the 
ambiguously defined variables like $\theta_\nu$. We will use both involutions: in some sense $\iota$ is simpler and more natural, but $\iota'$ is needed
for compatibility with \cite{ilt13}.

\subsection{Transformation from $s$ to $t$ channel}
\label{sec:st_transformation}
\begin{figure}[h!]
\begin{center}

\begin{tikzpicture}[decoration={markings,mark=at position 0.25 with {\arrow{Stealth}}}]
\draw(0,0) ellipse [x radius=3.5cm,y radius = 1.5cm];
\draw[fill=black,radius=0.07](-2,0) circle [radius=0.07]
(-1,0.4) circle[radius=0.07]
(1,0) circle[radius=0.07]
(2.5,0) circle[radius=0.07];

\node[inner sep=0] (z0) at (-2.5,-0.5){}; 
\node at (-2.6,-0.7){$z_0$};

\draw[postaction={decorate}] (z0) to[out=45,in=-135] (-1.9,-0.1) arc [radius=0.15cm,start angle=-45,end angle=135] to[out=-135,in=45] (z0);

\draw[postaction={decorate}] (z0) to[out=15,in=-135] (-0.9,0.3) arc [radius=0.15cm,start angle=-45,end angle=135] to[out=-135,in=15] (z0);

\draw[postaction={decorate}] (z0) to[out=0,in=-155] (1.1,-0.13) arc [radius=0.15cm,start angle=-65,end angle=115] to[out=-155,in=0] (z0);

\draw[postaction={decorate}] (z0) to[out=-15,in=-155] (2.6,-0.13) arc [radius=0.15cm,start angle=-65,end angle=115] to[out=-155,in=-15] (z0);

\draw[fill=black] (z0) circle[radius=0.05];

\node at (-2,0.4) {$0$};
\node at (-1,0.8) {$t$};
\node at (1,0.4) {$1$};
\node at (2.5,0.4) {$\infty$};

\node[draw,circle,inner sep=2] at (0,1.2){$z$};

\begin{scope}[xshift=8cm, yscale=-1,xscale=-1]

\node[inner sep=0] (z0) at (-2.5,-0.5){}; 
\node[inner sep=-3] (z1) at (1.5,0.5){}; 
\node at (-2.5,-0.75){$1-z_0$};

\draw[postaction={decorate}] (z0) to[out=45,in=-135] (-1.9,-0.1) arc [radius=0.15cm,start angle=-45,end angle=135] to[out=-135,in=45] (z0);

\draw[postaction={decorate}] (z0) to[out=15,in=-135] (-0.9,0.3) arc [radius=0.15cm,start angle=-45,end angle=135] to[out=-135,in=15] (z0);

\draw[postaction={decorate}] (z0) to[out=0,in=-155] (1.1,-0.13) arc [radius=0.15cm,start angle=-65,end angle=115] to[out=-155,in=0] (z0);

\draw[postaction={decorate}] (z0) to[out=-15,in=-155] (2.6,-0.13) arc [radius=0.15cm,start angle=-65,end angle=115] to[out=-155,in=-15] (z0);

\draw(0,0) ellipse [x radius=3.5cm,y radius = 1.5cm];
\draw[fill=black,radius=0.07](-2,0) circle [radius=0.07]
(-1,0.4) circle[radius=0.07]
(1,0) circle[radius=0.07]
(2.5,0) circle[radius=0.07];

\draw[fill=black] (z0) circle[radius=0.05];
\draw[fill=black] (z1) circle[radius=0.05];

\node at (-2,0.4) {$1$};
\node at (-1,0.8) {$1-t$};
\node at (1,0.4) {$0$};
\node at (2.5,0.4) {$-\infty$};

\node[draw,circle,inner sep=2] at (0,-1.2){$z$};

\draw[dashed,-Stealth] (z0) to[out=20,in=-120](-1.6,0.2) to[out=60,in=-90] (-1.6,0.8)  to[out=90,in=160] (1.4,0.55);

\end{scope}

\end{tikzpicture}

\end{center}
\caption{Transformation of loops under $z\mapsto 1-z$.}
\label{fig:1_minus_z}
\end{figure}
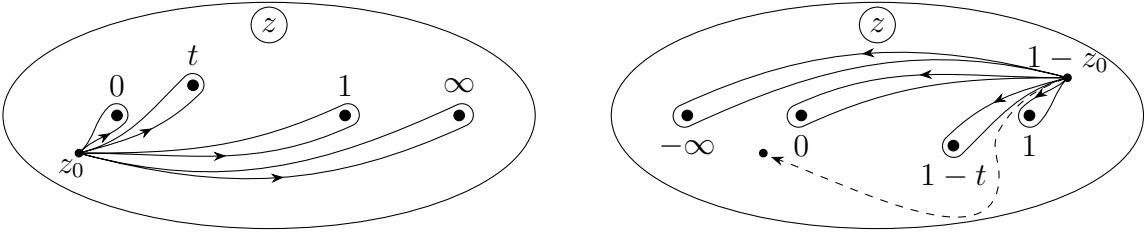

We review here how to connect the tau-functions $\tau_{01}(t)$ and $\tau_{1t}(t)$,   introduced in (\ref{eq:isoVira}) and in (\ref{eq:tau1t}). In order to do that we study here  the transformation of the loop $\gamma \in  \pi_1(\mathbb CP^1\setminus\{0,t,1,\infty\})$ under the map $z\mapsto 1-z$, $\gamma\to \tilde{\gamma}$. Figure~\ref{fig:1_minus_z} shows the transformation of the generating loops $\gamma_{\nu}$, $\nu=0,t,1,\infty$ under this map.  The corresponding action on the monodromy matrices $M_{\nu}$ will be denoted by $\tilde{M}_{\nu}$, $M_{\nu}\to \tilde{M}_{\nu}$, $\nu=0,t,1,\infty$.
Consider two monodromy points $P=(\bs\theta,\sigma_{0t},s_{0t})$ and $\tilde{P}=(\bs\theta,\sigma_{1t},s_{1t})$ ($P,\tilde{P}\in \monSpace$) that describe respectively the monodromies  of  $\tau_{01}(t)$ and $\tau_{1t}(t)$. We show here how $P$ and $\tilde{P}$ are related one to the other. 

By evoking the same considerations as the the ones discussed previously for the involution $\iota$, the transformed monodromy matrices can be written as
\eq{
\label{eq:transf_map}
\tilde M_0=M_tM_1M_t^{-1}\,,\quad \tilde M_t=M_t\,,\quad \tilde M_1=M_0\,,\quad \tilde M_\infty=M_0^{-1}M_\infty M_0\,.
}

As far the invariant quantities $\tilde p_{0t}=\tr \tilde M_0\tilde M_t$, $\tilde p_{1t}=\tr \tilde M_1\tilde M_t$ and  $\tilde p_{01}=\tr \tilde M_1\tilde M_\infty$ are concerned, formulas strictly analogous to (\ref{eq:p0t}) should hold. In particular one expects that the expressions for  $\tilde p_{0t},\tilde p_{1t}$ and $\tilde p_{01}$ are  obtained from \eqref{eq:p0t} by replacing $\theta_0\to\theta_1$, $\theta_0\to\theta_1$,
$\sigma_{0t}\to \sigma_{1t}$, $s_{0t}\to s_{1t}$. This is true except a sign subtlety: from (\ref{eq:transf_map}), $\tilde p_{0t}=p_{1t}$ and $\tilde p_{1t}=p_{0t}$, $\tilde p_{01}=\tr M_0M_t^{-1}M_1M_t=\iota(p_{01})$, where  $\iota(p_{01})$ is defined in (\ref{eq:iotap01}). Therefore $p_{01}=\iota(\tilde p_{01})$, which effectively  changes sign in the exponents. Using these results, we can express the quantities in $P$ in terms of the coordinates of $\tilde{P}$:
\eq{
p_{1t}=2\cos2\pi\sigma_{1t}\,,
\\
p_{0t}=\tr \tilde M_1\tilde M_t=\frac1{\sin^2 2\pi\sigma_{1t}}\lb\frac 12\lb p_tp_0+p_1p_\infty\rb-\frac14\lb p_0p_1+p_tp_\infty\rb p_{1t}-\right.\\
\left.- 4 \Sum_{\epsilon'=\pm}s_{1t}^{-\epsilon'}\Prod_{\epsilon=\pm}\sin \pi(\epsilon\theta_1+\theta_t+\epsilon'\sigma_{1t})\sin\pi(\epsilon\theta_\infty+\theta_0+\epsilon'\sigma_{1t})\rb\,,
\\
p_{01}=\iota\lb\tr \tilde M_0\tilde M_1\rb=
\frac1{\sin^2 2\pi\sigma_{1t}}\lb\frac12\lb p_0p_1+p_tp_\infty\rb-\frac14\lb p_tp_0+p_1p_\infty\rb p_{1t}+ \right.\\
\left.+ 4 \Sum_{\epsilon'=\pm}s_{1t}^{-\epsilon'} e^{-2\pi i\sigma_{1t}\epsilon'}\Prod_{\epsilon=\pm} \sin \pi(\epsilon\theta_1+\theta_t+\epsilon'\sigma_{1t})\sin\pi(\epsilon\theta_\infty+\theta_0+\epsilon'\sigma_{1t})\rb\,.
\label{eq:p1t}
}
Notice that (\ref{eq:p0t}) and (\ref{eq:p1t}) are symmetric under the sign inversion: $(\sigma_{0t},s_{0t})\mapsto (-\sigma_{0t},s_{0t}^{-1})$, and $(\sigma_{1t},s_{1t})\mapsto (-\sigma_{1t},s_{1t}^{-1})$.
This means that each pair of coordinates, $(\sigma_{0 1},s_{01})$ or $(\sigma_{1 1},s_{1t})$ actually realizes the two-fold cover of $\monSpaceC$:
\eq{
(\sigma_{\mu\nu},s_{\mu\nu})\simeq (-\sigma_{\mu\nu},s_{\mu\nu}^{-1})\,.
\label{eq:identification}
}

By comparing the (\ref{eq:p0t}) and (\ref{eq:p1t}), we can find the relation between the coordinates $(\sigma_{0t},s_{0t})$ and  $(\sigma_{1t},s_{1t})$, i.e. the relation between $P$ and $\tilde{P}$. We proceed like follows: first we use $p_{1t}$ in terms of $\sigma_{0t}$ and $s_{0t}$ from (\ref{eq:p0t}), thus fixing the value of $\sigma_{1t}=\sigma_{1t}(\sigma_{0t},s_{0t})$ using the expression for $p_{1t}$ in  (\ref{eq:p1t}). Finally, using the following formula
obtained from \eqref{eq:p1t}:
\eq{
s_{1t}^{\pm1}
=
\frac{\mp 2i\sin 2\pi\sigma_{1t}(p_{01}+e^{\mp 2\pi i\sigma_{1t}}p_{0t})
-e^{\mp 2\pi i\sigma_{1t}}(p_0p_1+p_tp_\infty)
+(p_tp_0+p_1p_\infty)
}{16\Prod_{\epsilon=\pm} \sin \pi(\epsilon\theta_1+\theta_t\mp\sigma_{1t})\sin\pi(\epsilon\theta_\infty+\theta_0\mp\sigma_{1t})}\,
\label{eq:s1t}
}
we can express  the dependence of $s_{1t}$ on $(\sigma_{0t},s_{0t})$.
\subsubsection{Examples}
\label{sec:st_examples}
As an example of application of these formulas we consider the case $\theta_0=\theta_t=\theta_1=\theta_\infty=\frac14$, which corresponds to
quasi-permutational monodromy. We introduce for convenience
\eq{
s_{\mu\nu}=-e^{2\pi i\eta_{\mu\nu}}\,.
\label{eq:eta}
}
One can check that the following equality follows from \eqref{eq:p0t}: $\cos 2\pi\sigma_{1t}=\cos 2\pi\eta_{0t}$. As a consequence we have either
$\sigma_{1t}=\eta_{0t}$ or $\sigma_{1t}=-\eta_{0t}$. Finally we get
\eq{
  (\sigma_{1t},\eta_{1t})=(\eta_{0t},-\sigma_{0t})\qquad \text{or}\qquad (\sigma_{1t},\eta_{1t})=(-\eta_{0t},\sigma_{0t}).
  \label{eq:at_st}
}
We easily see that transformations \eqref{eq:at_st} preserve symplectic form \eqref{eq:AtiyahBott}. This property holds also in the general case:
\eq{
d\omega=d\sigma_{0t}\wedge d\eta_{0t}=d\sigma_{1t}\wedge d\eta_{1t}\,.
}

In the general case transformation $(\sigma_{0t},\eta_{0t})\mapsto(\sigma_{1t},\eta_{1t})$ is quite complicated. However, for our purposes, see  Subsection~\ref{sec:Analytic}, we need to solve a simpler problem, namely the one of finding how this transformation changes the topological type of the cycle
\eq{
  \mc C_{h,r}=\{\sigma_{0t},\eta_{0t}|\operatorname{Im}\sigma_{0t}=h,\operatorname{Im}\eta_{0t}=r\}\,,
  \label{eq:Chr}
}
where we consider the limit~\footnote{For practical applications it is enough to take $A\gg  1$ and $A\gg \im \theta_\nu$.} $|h|>A, |r|>A, |r-h|>A$, $A \to \infty$. First take this limit in \eqref{eq:p0t}:
\eqs{
  &p_{0t}=e^{-2\pi i\sigma_{0t}\sign h}+O(e^{-2\pi A})\,,\\
  &p_{1t}=e^{-2\pi i(\eta_{0t}+\theta_t+\theta_1+\frac12)\sign r}+O(e^{-2\pi A})\,,\\
  &p_{01}=e^{-2\pi i(\eta_{0t}-\sigma_{0t}+\theta_t+\theta_1)\sign(r-h)}+O(e^{-2\pi A})\,.
  }
  Now one has to fix sign for $\sigma_{1t}$:
\eq{
  \sigma_{1t}=-(\eta_{0t}+\theta_t+\theta_1+\frac12)\sign r + O(e^{-2\pi A})\,.
  \label{eq:sigma1t}
}
Then compute the limit of $s_{1t}$ using \eqref{eq:s1t}:
\eq{
  s_{1t}^{\pm1}=\quad O(e^{-2\pi A})+\\ +\frac{\mp e^{-2\pi i(\eta_{0t}+\theta_t+\theta_1+\frac12)\sign r}(e^{-2\pi i(\eta_{0t}-\sigma_{0t}+\theta_t+\theta_1)\sign(r-h)}+ e^{\pm 2\pi i(\eta_{0t}+\theta_t+\theta_1+\frac12)\sign r -2\pi i \sigma_{0t} \sign h})}{e^{-4\pi i(\eta_{0t}+\theta_t+\theta_0)\sign r \mp 2\pi i(\theta_t+\theta_0)}}\,.
\label{eq:s1t_limit}
}
Now there are two possible situation, $\sign h\cdot\sign r>0$ and $\sign h\cdot\sign r<0$.
In the first situation only the second term in \eqref{eq:s1t_limit} dominates in the formula for $s_{1t}^{-1}$:
\eq{
s_{1t}^{-1}= e^{-2\pi i\sigma_{0t}\sign r +4\pi i(\theta_0-\theta_1)\sign r-2\pi i(\theta_t+\theta_0)}+O(e^{-2\pi A})\,,
}
while in the second situation only the first term in \eqref{eq:s1t_limit} dominates in the formula for $s_{1t}^{+1}$:
\eq{
s_{1t}^{+1} = e^{2\pi i\sigma_{0t}\sign r +4\pi i(\theta_0-\theta_1)\sign r+2\pi i(\theta_t+\theta_0)}+O(e^{-2\pi A})\,.
}
One can write these expressions in a unified way:
\eq{
  \eta_{1t}=\sigma_{0t}\sign r - 2(\theta_0-\theta_1)\sign h - (\theta_0+\theta_t-\frac12)+O(e^{-2\pi A})\,.
  \label{eq:eta1t}
}
Notice that these limiting transformations \eqref{eq:sigma1t} and \eqref{eq:eta1t} are very similar to \eqref{eq:at_st}, and they also transform contour $\mc{C}_{h,r}$ to contour $\mc{C}_{\tilde h,\tilde r}$ up to small perturbations of order $O(e^{-2\pi A})$, where
\eqs{
  &\tilde h=-|r|-(\im\theta_t+\im\theta_1)\sign r\,,\\
  &\tilde r=h\sign r -2(\im \theta_0-\im\theta_1)\sign h-\im\theta_0-\im\theta_t\,.
}
So at the level of homology classes we have transformation
\eq{
  \left[\mc{C}_{h,r}\right]\mapsto\left[\mc{C}_{\tilde h,\tilde r}\right]\,.
  \label{eq:homology_transformation}
}

\subsection{Braid group $\mathbb B_4$}
\label{sec:braid}

More in general we consider the action of the braid group $\mathbb B_4$ on the monodromy manifold. Indeed, the action of this group $\mathbb B_4$ on the tau-function performs the analytic continuation of the tau-function in the variable $t$ and different re-expansions
between different channels, like $s$- and $t$-channels.
\noindent  In Artin representation, and element $b$ of the braid group $\mathbb{B}_4$ is generated by a product of  three basic elements,  $\left(b_{0t},b_{t1},b_{1,\infty}\right)$. For instance the element $b_{0t}$ braids path passing through  $0$  below the path passing through $t$, as in this figure:
\begin{center}
\begin{tikzpicture}[scale=1,every node/.style={scale=1}]
\braid[number of strands=4] (braid) a_1;
\draw (0,-1) node[left]{$b_{0t}:$};
\draw (1.4,-2) node[left]{$0$};
\draw (2.2,-2) node[left]{$t$};
\draw (3.5,-2) node[left]{$1$};
\draw (4.5,-2) node[left]{$\infty$};
\end{tikzpicture}
\end{center}
Studying the corresponding transformation on the loops, one can derive the action of the generators $(b_{0t},b_{t1},b_{1,\infty})$ on the matrices $M_{\nu}$:  
\eq{
b_{0t}: (M_0,M_t,M_1,M_\infty)\mapsto (M_t,M_t^{-1}M_0M_t,M_1,M_\infty)\\
b_{t1}: (M_0,M_t,M_1,M_\infty)\mapsto (M_0,M_1,M_1^{-1}M_tM_1,M_\infty)\,\\
b_{1\infty}: (M_0,M_t,M_1,M_\infty)\mapsto (M_0,M_t,M_\infty,M_\infty^{-1}M_{1}M_{\infty})\,.
\label{eq:braidaction}
}

Let us consider for instance the element $b=b_{0t}^2$:
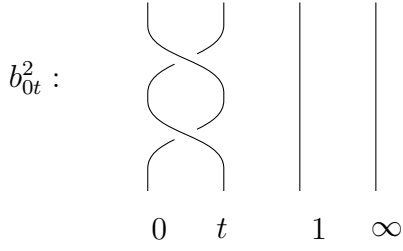
\begin{center}
\begin{tikzpicture}[scale=1,every node/.style={scale=1}]
\braid[number of strands=4] (braid) a_1 a_1;
\draw (0,-1) node[left]{$b_{0t}^2:$};
\draw (1.4,-3) node[left]{$0$};
\draw (2.2,-3) node[left]{$t$};
\draw (3.5,-3) node[left]{$1$};
\draw (4.5,-3) node[left]{$\infty$};
\end{tikzpicture}
\end{center}
Under the action of $b_{0t}^2$, the monodromy matrices transform as:
\eq{
b_{0t}^2:(M_0,M_t,M_1,M_\infty)\mapsto (M_t^{-1}M_{0}M_{t},M_t^{-1}M_0^{-1}M_{t}M_0M_{t},M_1,M_\infty),
}
that, from (\ref{eq:M0Mt}) and (\ref{eq:Mt}), implies:
\eq{
\label{eq:b0t2}
b_{0t}^2\left(\bs \theta, \sigma_{0t}, s_{0t}\right)= \left( \bs \theta, \sigma_{0t}, s_{0t}e^{4 \pi i \sigma_{0t}}\right)
}
Finally, one can verify that the $s-$channel to $t-$channel transformation (\ref{eq:transf_map}) corresponds to the element $b_{st}\in\mathbb B_4$, 
\eq{
b_{st}= b_{t1}b_{0t}b_{t1},
\label{eq:bst}
} associated to the following braid :
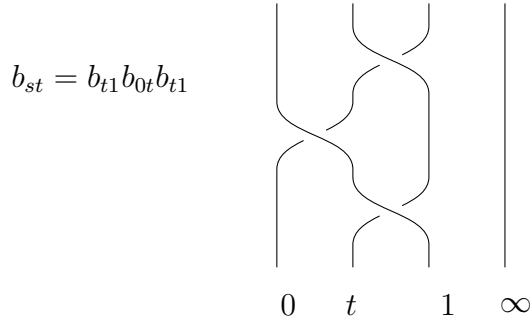
\begin{center}
\begin{tikzpicture}[scale=1,every node/.style={scale=1}]
\braid[number of strands=4] (braid) a_2 a_1 a_2;
\draw (0,-1) node[left]{$b_{st}=b_{t1}b_{0t}b_{t1}$};
\draw (1.4,-4) node[left]{$0$};
\draw (2.2,-4) node[left]{$t$};
\draw (3.5,-4) node[left]{$1$};
\draw (4.5,-4) node[left]{$\infty$};
\end{tikzpicture}
\end{center}
We have that:
\eq{
\label{eq:actionbst}
b_{st}\left((\bs \theta, \sigma_{0t}, s_{0t})\right)=(\tilde{\bs\theta}, \sigma_{1t}, s_{1t})
}
where $ \tilde{\bs\theta}=\left(\theta_1,\theta_t,\theta_0,\theta_{\infty}\right)$.

Let us now explain the relation between our description of crossing (fusion) and braiding transformations and the Moore--Seiberg formalism \cite{moore1989}.
In the case of general multi-point conformal blocks one has the action of the Moore-Seiberg groupoid generated by the two transformations: braiding, which moves two colliding fields around each other, and fusion, which performs local $s$--$t$ channel transformation of conformal blocks, or changes pants decomposition.
In general, fusion changes topological type of the pants decomposition.
However, it turns out that for 4 and 5 points all pants decompositions are the same up to permutation of the external states (for 6 points we already have two inequivalent pictures).
This means that one can fix some particular pants decomposition, for which $t$ collides with $0$ and $1$ collides with $\infty$~\footnote{At the level of braids this means that we add some extra structure describing pants decomposition. Here we may say that strands are divided into two groups, $(0, t)$ and $(1, \infty)$.}.
Then we relate all other decompositions to this one by the action of the braid group permuting external states and acting on the complex modulus $t$ by analytic continuation.
So for the 4-point case we can consider $\mathbb{B}_4$ instead of the Moore-Seiberg groupoid: for example, see Figure~\ref{fig:fusion-braiding} for the action of $b_{st}$ on 4-point conformal block.

\begin{figure}[h!]
\begin{center}
\begin{tikzpicture}

\begin{scope}[scale=0.9]
  \begin{scope}
    \draw(0,0)--(4,0) (1,0)--(1,2) (3,0)--(3,2);

    \node at(0.2,-0.25){$0$};
    \node at(0.2,0.3){$\theta_0$};

    \node at(3.8,0.25){$\infty$};
    \node at(3.8,-0.3){$\theta_\infty$};

    \node at(0.8,1.7){$t$};
    \node at(1.25,1.7){$\theta_t$};

    \node at(2.8,1.7){$1$};
    \node at(3.25,1.7){$\theta_1$};
    
    \node at (2,0.25){$\sigma_{0t}$};

  \end{scope}

  \begin{scope}[xshift=8cm]
    \draw(0,0)--(4,0) (1,0)--(1,1) (3,0)--(3,1);
    \draw(0,0) to[out=180, in=-120] (0,1);
    \draw (0,1) to [out=60,in=-90](3,3);
    \draw[fill=white,radius=0.15cm,white](1,1.73) circle;
    \draw (1,1.3) -- (1,3);
    \draw[fill=white,radius=0.15cm,white](-0.18,0.5) circle;
    \draw(3,1) to[out=90,in=0] (-2,0);

    \node at(-1.8,-0.25){$1$};
    \node at(-1.8,0.3){$\theta_0$};

    \node at(3.8,0.25){$\infty$};
    \node at(3.8,-0.3){$\theta_\infty$};

    \node at(0.4,2.7){$1-t$};
    \node at(1.25,2.7){$\theta_t$};

    \node at(2.75,2.8){$0$};
    \node at(3.25,2.7){$\theta_1$};

    \node at (2,0.25){$\sigma_{1t}$};

  \end{scope}

  \begin{scope}[xshift=13cm]
    \draw(0,0)--(4,0) (2,0)--(2,1);
    \draw (2,1)--(1,2) (2,1)--(3,2);

    \node at(0.2,-0.25){$1$};
    \node at(0.2,0.3){$\theta_0$};

    \node at(3.8,0.25){$\infty$};
    \node at(3.8,-0.3){$\theta_\infty$};

    \node at(0.55,1.7){$1-t$};
    \node at(1.5,1.9){$\theta_t$};

    \node at(2.7,2){$0$};
    \node at(3,1.6){$\theta_1$};

    \node at (2.4,0.5){$\sigma_{1t}$};

  \end{scope}

\node at(4.95,0){$\longmapsto$};
\node at(4.95,0.4){$b_{st}$};
\node at(12.45,0){$=$};

  \begin{scope}[xshift=13cm, yshift=-4cm]
    \draw(0,0)--(4,0) (1,0)--(1,2) (3,0)--(3,2);

    \node at(0.2,-0.25){$0$};
    \node at(0.2,0.3){$\theta_1$};

    \node at(3.8,0.25){$\infty$};
    \node at(3.8,-0.3){$\theta_\infty$};

    \node at(0.45,1.7){$1-t$};
    \node at(1.25,1.7){$\theta_t$};

    \node at(2.8,1.7){$1$};
    \node at(3.25,1.7){$\theta_0$};
    
    \node at (2,0.25){$\sigma_{1t}$};

  \end{scope}

\node[rotate=90] at(15,-1){$=$};

\node at(11.95,-4){$=$};

\node at(4,-4){$\int_{\mathbb R+i\Lambda}d\sigma_{0t}F\left[\begin{array}{cc}\theta_0& \theta_t\\ \theta_\infty & \theta_1\end{array} ; \begin{array}{c}\sigma_{0t}\\\sigma_{1t}\end{array}\right] \times $};

  \begin{scope}[xshift=7cm,yshift=-4cm]
    \draw(0,0)--(4,0) (1,0)--(1,2) (3,0)--(3,2);

    \node at(0.2,-0.25){$0$};
    \node at(0.2,0.3){$\theta_0$};

    \node at(3.8,0.25){$\infty$};
    \node at(3.8,-0.3){$\theta_\infty$};

    \node at(0.8,1.7){$t$};
    \node at(1.25,1.7){$\theta_t$};

    \node at(2.8,1.7){$1$};
    \node at(3.25,1.7){$\theta_1$};
    
    \node at (2,0.25){$\sigma_{0t}$};

  \end{scope}

\end{scope}
  
\end{tikzpicture}
\end{center}
\caption{Fusion transformation as the action of the braid group.}
\label{fig:fusion-braiding}
\end{figure}

The most important and non-trivial element is $b_{st}$, but there are also many elements which do not change the channel of expansion.
They give rise to nice non-trivial relations between conformal blocks that can be found in \cite{gil12}. For example, there is an identity for $b_{1\infty}$:
\eq{
\mc B\left(\theta_0,\theta_t,\theta_1,\theta_\infty;\sigma_{0t};\frac{t}{t-1}\right)=
(-1)^{\sigma_{0t}^2-\theta_0^2-\theta_t^2}(1-t)^{2\theta_t^2}
\mc B\left(\theta_0,\theta_t,\theta_\infty,\theta_1;\sigma_{0t};t\right)\,.
\label{eq:half_brading}
}
Here $-1$ should be treated either as $e^{\pi i}$ or as $e^{-\pi i}$, dependently on the direction of analytic continuation, and comes from the power of $t$ in the normalization factor.

We also have the same simple formula for $b_{0t}b_{1\infty}^{-1}$:
\eq{
\mc B\left(\theta_t,\theta_0,\theta_\infty,\theta_1;\sigma_{0t};t\right)=
(1-t)^{\theta_0^2+\theta_\infty^2-\theta_1^2-\theta_t^2}\mc B\left(\theta_0,\theta_t,\theta_1,\theta_\infty;\sigma_{0t};t\right)\,.
\label{eq:double_half_brading}
}
Together these three elements $b_{st}, b_{0t}, b_{1\infty}$ generate the whole braid group. Notice also that since this braid group acts on sphere, we have extra relation $b_{0t}^2=b_{1\infty}^2$, coming from the overall coordinate rotation $z\mapsto z e^{2\pi i}$, so effectively this group almost reduces to $\mathbb B_3$.
\section{Crossing symmetric correlation function from tau functions}

It is convenient to introduce the  tau-function $\hat{\tau}_{0t}$ that differs from the \eqref{eq:isoVira} only by a normalization:
\eq{
\hat{\tau}_{0t}(P;t)= \sum_{n \in \mathbb{Z}} s_{0t}^n \frac{C(\bs \theta;\sigma_{0t}+n)}{C(\bs \theta;\sigma_{0t})} \mc B(\bs \theta;\sigma_{0t}+n;t),\quad P=(\bs \theta,\sigma_{0t},s_{0t})
\label{taus}}

A CFT correlation function  is obtained by gluing the holomorphic  and anti-holomorphic conformal block.  We have shown before the transformation of the point $P$ under the conjugation $z\mapsto \bar{z}$. Using \eqref{eq:iotaP1} and \eqref{eq:iotaP}, the corresponding tau function can be written in two different forms:
\eq{
\hat{\tau}_{0t}(\iota'(P),\bar{t})=\sum_{n \in \mathbb{Z}} \left(\iota'(s_{0t})\right)^n \frac{C(\bs \theta;\sigma_{0t}+n)}{C(\bs \theta;\sigma_{0t})}\; \mc B(\bs \theta;\sigma_{0t}+n;\bar{t})\,,\\
\hat{\tau}_{0t}(\iota(P),\bar{t})=\sum_{n \in \mathbb{Z}} s_{0t}^{-n} \frac{C(-\bs \theta;\sigma_{0t}+n)}{C(-\bs \theta;\sigma_{0t})}\; \mc B(\bs \theta;\sigma_{0t}+n;t)\,.
}
then, using \eqref{eq:C_GIL}, \eqref{eq:involution} and the following property of the  Barnes functions
\eq{
\frac{G(1-\nu+n)}{G(1-\nu)}=(-1)^{\frac{n(n-1)}2}\frac{G(1+\nu-n)}{G(1+\nu)}\lb\frac{\pi}{\sin\pi\nu}\rb^n\,,
}
we can show that these two functions coincide: $\hat{\tau}_{0t}(\iota'(P),\bar{t})=\hat{\tau}_{0t}(\iota(P),\bar{t})$.

Taking into account the identity:
\eq{
C(\bs \theta,\sigma_{0t})C(-\bs \theta,\sigma_{0t})=\Phi(\theta_0,\theta_t,\sigma_{0t})\Phi(\sigma_{0t},\theta_1,\theta_\infty),
\label{eq:C_Phi}
}
where $\Phi$ has been introduced in (\ref{eq:Phidef}), we define the non-holomorphic function $\taut$ as:
\eq{
\taut(P,t,\bar t)=\tau_{0t}(P,t)\tau_{0t}(\iota(P),\bar t)=\Phi(\theta_0,\theta_t,\sigma_{0t})\Phi(\sigma_{0t},\theta_1,\theta_\infty)\hat\tau_{0t}(P,t)\hat\tau_{0t}(\iota'(P),\bar t)\,.
\label{eq:taut}
}
One of the good properties of $\taut(P,t,\bar t)$ is that it is an actual function on the monodromy manifold $\monSpaceC$, while $\tau_{0t}(P,t)$ and 
$\tau_{0t}(\iota(P),t)$ are only sections of some line bundles.
Namely, their periodicity properties are given by $\tau_{0t}(\bs\theta,\sigma_{0t}+1,s_{0t};t)=
s_{0t}^{-1}\tau_{0t}(\bs\theta,\sigma_{0t},s_{0t};t)$ and $\tau_{0t}(-\bs\theta,\sigma_{0t}+1,s_{0t}^{-1};t)=
s_{0t}\tau_{0t}(-\bs\theta,\sigma_{0t},s_{0t}^{-1};t)$, whereas functions on $\monSpaceC$ are periodic under such shift.

\subsection{CFT correlation function in terms of $\monSpace$ averages}
Using  the  $\taut(P,t,\bar t)$ function, that contains a combination of holomorphic and anti-holomorphic conformal blocs, and the properties of the manifold $\monSpace$,  we want to define a function $\mc F_{\bs \theta}(t,\bar{t})$ that satisfies the following properties:
\begin{itemize}
\item To be single-valued in the variable $t$.
\item To be crossing invariant, i.e. the two expressions, in terms of the $s-$channel and in terms of the  $t-$channel conformal blocks, are equivalent.
\end{itemize}

Let consider a general  distribution $d\mu_{\bs \theta}(P)$ on $\monSpace$ that is supported by $\monSpaceC$, i.e.:
\eq{
d\mu_{\bs \theta}(P) = 0 \quad \text{for}\; P=(\bs \theta',\sigma_{0t},s_{0t}) \;\text{and}\; \bs \theta'\neq \bs \theta.
}
We will focus on the function $\mc F(t,\bar{t})[d\mu_{\bs\theta}]$, that is defined as the $ d\mu_{\bs \theta}$ average of the function  $\taut(P,t,\bar t)$ \eqref{eq:taut} over the space $\monSpace$:
\eq{
\mc F(t,\bar{t})[d\mu_{\bs\theta}]=\Int_\monSpace d\mu_{\bs\theta}(P)\;\taut(P,t,\bar t).
\label{eq:F}
}
The main point of this construction is that the crossing invariance of   $\mc F(t,\bar{t})[d\mu_{\bs\theta}]$ can be assured by properly choosing the distribution $d\mu_{\bs\theta}$. 

\subsection{Weakly equivalent and weakly invariant distributions}
To be more concrete, let us first introduce the notion of weak equivalence  between distributions on $\monSpaceC$. We say that two distributions $d\mu_1(P)$ and $d\mu_2(P)$ are weakly equivalent if their integrals  with all reasonable~\footnote{To be rigorous, one should say that by reasonable functions we mean the space of $\taut$ functions at all possible $t$'s.
This definition seems to be not very useful for classificational problems, so probably there might be some simpler description: for example, functions whose singularities in the finite domain are only $\sigma_{0t}=0,\frac12$. We don't actually know what is the best definition, since we are using these distributions only to average tau functions.} meromorphic test functions $f(P)$ coincide:
\eq{
d\mu_1\simeq d\mu_2\Leftrightarrow \forall f: \int d\mu_1(P) f(P)=\int d\mu_2(P) f(P)\,.
\label{eq:weak_equivalence}
}
We can now define a ``weakly invariant'' distribution. Consider an element $b$ of the four strands braid group $\mathbb B_4$, $b\in \mathbb B_4$. A distribution $\mu_{\bs\theta}(P)$ is weakly  invariant if the transformed distribution, $b\lb d\mu_{\bs\theta}(P)\rb$~\footnote{By transformed distribution we denote the inverse of the pull-back:$$\forall f: \int b(d\mu(P)) f(b(P))=\int d\mu(P) f(P).$$
  Sometimes we also use another notation: $d\mu(b(P))\equiv b(d\mu(P))$.}  is  weakly equivalent to the distribution with transformed $\bs\theta$:
\eq{
b\lb d\mu_{\bs\theta}(P)\rb\simeq d\mu_{b(\bs\theta)}(P)
\label{eq:weak_invariant}
}
The action of the element $b$ on $\bs \theta$ is simply a permutation of indexes. On the other hand, the action of the braid group on  the distribution is highly non-trivial as the transformation of the point $P\to b(P)$ is in general quite complicated, as we have seen in the previous sections for some special braid action.

\subsection{Weakly invariant distribution and crossing invariance}

If the distribution $d\mu_{\bs\theta}(P)$ is weakly invariant under $\mathbb B_4$ action, then the function $\mc F(t,\bar{t})[d\mu_{\bs\theta}]$, defined by (\ref{eq:F}), provides a crossing-invariant  CFT correlation function $\mc F_{\bs \theta}(t,\bar{t})= \mc F(t,\bar{t})[d\mu_{\bs \theta}]$. 
In particular we show  that  the following theorem holds:
\begin{theorem}
  \label{thm1}
  Consider a weakly invariant distribution $d\mu_{\bs\theta}(P)$. The function $\mc F(t,\bar{t})[d\mu_{\bs\theta}]$, defined by \eqref{eq:F}, is a single-valued function of $t$ and satisfies the properties:
\begin{align}
&A)\; \mc F(e^{2\pi i} t,e^{-2\pi i}\bar{t})[d\mu_{\bs \theta}]\;=\;\mc F(t,\bar{t})[d\mu_{\bs \theta}], \label{singlevalue} \\
&B)\; \mc F(t,\bar{t})[d\mu_{\bs \theta}]=\mc F(1-t,1-\bar{t})[d\mu_{\tilde{\bs \theta}}]\, 
\nonumber \\
&\text{with:}\;\bs \theta= \left(\theta_0,\theta_t,\theta_1,\theta_\infty\right),\quad \tilde{\bs \theta}=\left(\theta_1,\theta_t,\theta_0,\theta_\infty\right)
\label{crossing}
\end{align}
\end{theorem}
The above theorem implies the crossing invariance of the CFT correlation function~\footnote{We might also claim the stronger invariance under the half-rotation $t\mapsto e^{\pi i} \frac{t}{1-t}$ that permutes $\theta_0$ with $\theta_t$ or $\theta_1$ with $\theta_\infty$, \eqref{eq:half_brading}, \eqref{eq:double_half_brading}, but for simplicity we do not focus on such finite symmetries.}.

{\bf Proof.} We start by considering property $A$, see (\ref{singlevalue}). This property is related to the transformation of the correlation function under the double braid  action $b_{0t}^2$ that implements the rotation of  the position $t$ around $0$. The action of $b_{0t}^2$ on $\monSpace$ is given by (\ref{eq:b0t2}). One verifies that 
\eq{
\taut(\bs\theta;\sigma_{0t},s_{0t};e^{2\pi i}t,e^{-2\pi i}t)=\taut(\bs\theta;\sigma_{0t},s_{0t}\cdot e^{4\pi i\sigma_{0t}};t,\bar t).\label{eq:ttbraiding}
}
This is the consequence of two identities:
\eq{
\tau_{0t}(\bs\theta;\sigma_{0t},s_{0t};e^{2\pi i}t)=e^{2\pi i(\sigma_{0t}^2-\theta_0^2-\theta_t^2)}
\tau_{0t}(\bs\theta;\sigma_{0t},s_{0t}\cdot e^{4\pi i\sigma_{0t}};t)\,,\\
\tau_{0t}(-\bs\theta;\sigma_{0t},s_{0t}^{-1};e^{-2\pi i}\bar t)=e^{-2\pi i(\sigma_{0t}^2-\theta_0^2-\theta_t^2)}
\tau_{0t}(-\bs\theta;\sigma_{0t},s_{0t}^{-1}\cdot e^{-4\pi i\sigma_{0t}};\bar t)\,.
}
 One can study the behavior of correlation function under such analytic continuation
\eq{
\mc F(e^{2\pi i}t, e^{-2\pi i} \bar{t})[d\mu_{\bs\theta}]=\Int_\monSpace d\mu_{\bs\theta}(P)\;\taut(P,e^{2\pi i}t,e^{-2\pi i}\bar t)=\Int_\monSpace d\mu_{\bs\theta}(P)\;\taut(b_{0t}^2(P),t)=\\=
\Int_\monSpace d\mu_{\bs\theta}(b_{0t}^2(P))\;\taut(b_{0t}^2(P),t)=\Int_\monSpace d\mu_{\bs\theta}(\tilde P)\;\taut(\tilde P,t)=\mc F(t, \bar{t})[d\mu_{\bs\theta}]
\label{proof1}}
In the above series of identities we have used \eqref{eq:ttbraiding} and the weak invariance of $d\mu_{\bs\theta}$ that allows  to replace $d\mu_{\bs \theta}(P)\to d\mu_{\bs \theta}(b_{0t}^2(P))$. Notice that if we  set  $d\mu_{\bs\theta}(P)$ to be given by the closed differential $2$-form (in Darboux coordinates):
\eq{
\label{eq:Darboux}
d\mu_{\bs\theta}(P) = d\sigma_{0t} \wedge \frac{d s_{0t}}{s_{0t}},
}
one can directly verify that the distribution is invariant under the change of coordinates $(\bs \theta,\sigma_{0t},s_{0t})\to(\bs \theta,\sigma_{0t},s_{0t}e^{4\pi i \sigma_{0t}})$,  $d\mu_{\bs \theta}(P)= d\mu_{\bs \theta}(b_{0t}^2(P))$.

We give a proof now of property $B$, see (\ref{crossing}). In this case we consider the braid element $b_{st}$ defined by (\ref{eq:bst}). The action of $b_{st}$ on $\monSpace$ is given by (\ref{eq:actionbst}).
We prove below that:
\eq{
\label{eq:identityst}
\taut(\theta_1,\theta_t,\theta_0,\theta_\infty;\sigma_{1t},s_{1t};1-t)=
\taut(\theta_0,\theta_t,\theta_1,\theta_\infty;\sigma_{0t},s_{0t};t)\,.
}
The above identity, together with (\ref{eq:weak_invariant}), implies the property $B$:
\eq{
\mc F(t, \bar{t})[d\mu_{\bs\theta}]=\Int_\monSpace d\mu_{\bs\theta}(P)\;\taut(P,t,\bar t)=
\Int_\monSpace d\mu_{\bs\theta}(P)\;\taut(b_{st}(P),1-t,1-\bar t)=\\=
\Int_\monSpace b_{st}(d\mu_{\tilde{\bs\theta}}(P))\;\taut(b_{st}(P),1-t,1-\bar t)=
\Int_\monSpace d\mu_{\tilde{\bs\theta}}(b_{st}(P))\;\taut(b_{st}(P),1-t,1-\bar t)=\\=\mc F(1-t, 1-\bar{t})[d\mu_{\tilde{\bs\theta}}]
\label{proof2}}

To show (\ref{eq:identityst}), we use the definition \eqref{eq:taut}
\eq{
\frac{\taut(\bs\theta;\sigma_{0t},s_{0t};t,\bar t)}{
\taut(\ttheta;\sigma_{1t},s_{1t};1-t,1-\bar t)}
=
\frac{\Phi(\theta_0,\theta_t,\sigma_{0t})\Phi(\sigma_{0t},\theta_1,\theta_\infty)
}{\Phi(\theta_1,\theta_t,\sigma_{1t})\Phi(\sigma_{1t},\theta_0,\theta_\infty)}\frac{\hat\tau_{0t}(\bs\theta;\sigma_{0t},s_{0t};t)
\hat\tau_{0t}(\bs\theta;\sigma_{0t},\iota'(s_{0t});\bar t)}{
\hat\tau_{0t}(\ttheta;\sigma_{1t},s_{1t};1-t)
\hat\tau_{0t}(\ttheta;\sigma_{1t},\iota'(s_{1t});1-\bar t)}
=
\\
=
\frac{\Phi(\theta_0,\theta_t,\sigma_{0t})\Phi(\sigma_{0t},\theta_1,\theta_\infty)
}{\Phi(\theta_1,\theta_t,\sigma_{1t})\Phi(\sigma_{1t},\theta_0,\theta_\infty)}
\bar\chi_{01}(\bs\theta;\sigma_{0t},\sigma_{1t};p_{01})\bar\chi_{01}(\bs\theta;\sigma_{0t},\sigma_{1t};\iota'(p_{01}))=1\,.
}
During this computation we first used \cite[formula 3.12]{ilt13} that defines connection constant:
\eq{
\hat\tau_{0t}(\bs\theta;\sigma_{0t},s_{0t};t)=\bar\chi_{01}(\bs\theta;\sigma_{0t},\sigma_{1t};p_{01})\hat\tau_{0t}(\ttheta;\sigma_{1t},s_{1t};1-t)\,,
}
and the most important relation \cite[formula 4.9]{ilt13}:
\eq{
\bar\chi_{01}(\bs\theta;\sigma_{0t},\sigma_{1t};p_{01})\bar\chi_{01}(\bs\theta;\sigma_{0t},\sigma_{1t};\iota'(p_{01}))
=
\frac{\Phi(\theta_1,\theta_t,\sigma_{1t})\Phi(\sigma_{1t},\theta_0,\theta_\infty)}{\Phi(\theta_0,\theta_t,\sigma_{0t})\Phi(\sigma_{0t},\theta_1,\theta_\infty)}\,.
\label{eq:chi_product}
}
The latter relation is quite surprising, since connection constant for single tau function $\bar\chi_{0t}$ is quite involved function of monodromy parameters,
but it turns out that the product of two connection constants simplifies drastically, and we need only this simple part~\footnote{
It is witten in \cite{ilt13} ``{\it A conceptual explanation of this intriguing coincidence is yet to be found }'', and we think that the application,
which was found here, indicates that this coincidence is not accidental. We also expect it to hold in more general cases.}.

\emph{So now we can state that the problem of construction of the crossing-invariant functions with $c=1$ given by the ansatz \eqref{eq:F}
is reduced to the problem of construction of weakly invariant distributions on $\monSpace$.}

\section{Examples of weakly invariant distributions}

We do not have yet a criteria for classifying completely neither the  weak equivalent  distributions, nor the weak invariant ones.  On the other hand we know some special cases that are shown below to provide non-trivial CFT solutions. These cases are:
\begin{enumerate}
\item The distributions that are invariant,  $b(d\mu_{\bs \theta}(P))=\mu_{b(\bs \theta)}(P)$. These distributions are obviously weakly invariant. As examples of invariant distribution, we may take either uniformly distributed measure on some invariant subset, or, more generally, invariant measures on invariant submanifolds.
\item The distributions that are given by an invariant holomorphic 2-form concentrated on two-dimensional submanifolds $M\subset\monSpace$, such that their homology classes $[M] \in H_2\lb\monSpace,\mathbb Z\rb$ are invariant under
the braid group action $\left[b(M)\right]=\left[M\right]$.
\end{enumerate}

Below we show that, among the cases mentioned above, we can find important CFT solutions.

\subsection{The Gaussian free field}
In \cite{gil12} it was observed that the Riccati solutions of the Painlev\'e VI were related to the conformal blocks appearing in the GFF.
The parameters $\bs \theta$ are given by:
\eq{
\bs \theta_{Riccati} = \left(\theta_0,\theta_t,\theta_1,-\theta_0-\theta_t-\theta_1\right)\,.
\label{eq: Riccati}
}
The remaining two variables of the monodromy data, that we can choose to be $(\sigma_{0t},\sigma_{1t})$ are also constrained to be:
\eq{
\sigma_{0t}= \theta_{0}+\theta_t, \quad \sigma_{1t}=\theta_{t}+\theta_{1}\,.
}
The corresponding tau function is simply given by \cite{gil12}:
\eq{
\tau_{0t}(P_{Riccati},t) = \mc B(\bs \theta_{Riccati},\theta_{0}+\theta_t, t) = t^{2 \theta_0 \theta_t} (1-t)^{2 \theta_1 \theta_t}\,.
}
It is then clear that by taking as measure:
\eq{
d \mu^{Riccati}_{\bs \theta_{Riccati}}(P)= \delta(\sigma_{0t}-\theta_{0}-\theta_t) \delta(\sigma_{1t}-\theta_{1}-\theta_t), 
\label{eq:distr_gff}
}
and using (\ref{eq:corr_GFF_fact}) and (\ref{eq:involution}), one gets:
\eq{
\mathcal{F}_{\bs \theta_{Riccati}}(t,\bar{t})=\mathcal{F}(t,\bar{t})[d \mu^{Riccati}_{\bs \theta_{Riccati}}]\,.
}
\subsection{The Runkel-Watts theory}
We reviewed this theory in section (\ref{sec:RW}). Now we show how this theory can be found in our approach.
Let us consider the real two-dimensional submanifold
\eq{
\monSpaceR\subset \monSpaceC
}
which corresponds to unitary monodromies. This manifold is necessarily invariant with respect to any transformation since the  braid transformation (\ref{eq:braidaction}) maps unitary monodromies to unitary ones. Therefore we can construct invariant distribution by restriction of invariant symplectic form
to $\monSpaceR$ using \eqref{eq:unitarity} and \eqref{eq:Runitary}
\eq{
d\mu^{unitary}_{\bs\theta}(P)=\frac1{2\pi i}d\sigma_{0t}\wedge\frac{ds_{0t}}{s_{0t}}\times\\\times
\delta(\operatorname{Im} \sigma_{0t})\delta(|s_{0t}|^2-R^{unitary}(\sigma_{0t},\bs\theta))RW(\theta_0,\theta_t,\sigma_{0t}) RW(\theta_1,\theta_\infty,-\sigma_{0t})\,.
\label{eq:dmu_unitary}
}
The Theorem \ref{thm1} assures the function $\mc F(t,\bar{t})[d\mu^{unitary}_{\bs\theta}]$, defined in (\ref{eq:F}), to be single-valued and crossing invariant. We show now that this function corresponds to the  correlation function $\mc F^{RW}_{\bs\theta}(t,\bar{t})$ defined in (\ref{eq:RW_correlator}):
\eq{
\label{eq:RWinv}
\mc F(t,\bar{t})[d\mu^{unitary}_{\bs\theta}]=\mc F^{RW}_{\bs\theta}(t,\bar{t})\,.
}
In order to compute the integral \eqref{eq:F} we introduce the short-hand notation
\eq{
\widetilde{R}(\sigma_{0t})=R^{unitary}(\sigma_{0t},\bs\theta)\,,\qquad
\widetilde{RW}(\sigma_{0t})=RW(\theta_0,\theta_t,\sigma_{0t})RW(\theta_\infty,\theta_1,\sigma_{0t})\,.
\label{eq:shorthand}
}
First we take the integral over $s_{0t}$:
\eq{
\mc F(t,\bar{t})[d\mu^{unitary}_{\bs\theta}]=\int_{0}^1 d\sigma_{0t} \Int_{|s_{0t}|^2=\widetilde{R}(\sigma_{0t})}
\frac{ds_{0t}}{2\pi is_{0t}} \widetilde{RW}(\sigma_{0t})\tau_{0t}(P,t)\tau_{0t}(\iota(P),\bar t)
=\\=
\int_{0}^1 d\sigma_{0t} \widetilde{RW}(\sigma_{0t}) \Int_{|s_{0t}|^2=\widetilde{R}(\sigma_{0t})}
\frac{ds_{0t}}{2\pi is_{0t}} \Sum_{n,m\in\mathbb Z}s_{0t}^{n-m}C(\bs\theta,\sigma_{0t}+n)C(-\bs\theta,\sigma_{0t}+m)\times\\
\times \mc B(\bs\theta,\sigma_{0t}+n,t)\mc B(-\bs\theta,\sigma_{0t}+m,\bar t)
=\\=
\int_{0}^1 d\sigma_{0t} \Sum_{n\in\mathbb Z}\widetilde{RW}(\sigma_{0t}+n)
C(\bs\theta,\sigma_{0t}+n)C(-\bs\theta,\sigma_{0t}+n)|\mc B(\bs\theta,\sigma_{0t}+n,t)|^2
\label{eq:int_s}
}
Now it remains to combine the sum into the integral over the real line and use \eqref{eq:Phidef}, \eqref{eq:C_Phi} and \eqref{eq:shorthand}:
\eq{
\mc F(t,\bar{t})[d\mu^{unitary}_{\bs\theta}]=\int_{-\infty}^\infty d\sigma_{0t} RW(\theta_0,\theta_t,\sigma_{0t})
RW(\theta_\infty,\theta_1,\sigma_{0t})\times\\\times
\time \times
\Phi(\theta_0,\theta_t,\sigma_{0t})\Phi(\theta_\infty,\theta_1,\sigma_{0t})\mc |\mc B(\bs\theta,\sigma_{0t},t)|^2\,.
}
Since the above expression reproduces the correlation function in the Runkel-Watts theory \eqref{eq:RW_correlator}, the identity (\ref{eq:RWinv}) is proven.

\subsection{Ashkin-Teller model}

Let us consider now the submanifold  $\mathcal{M}_{0,4}^{SU(2)}(\bs \theta_{Picard})$ generated by the following matrices
\eq{
M_0=\bpm 0&ie^{2\pi i\sigma_{0t}}\\ie^{-2\pi i\sigma_{0t}}&0\epm,\quad M_t=\bpm0&-i\\-i&0\epm,\quad
M_1=\bpm 0&ie^{-2\pi i\sigma_{1t}}\\ie^{2\pi i\sigma_{1t}}&0\epm
}
The above matrices, that have all eigenvalues $e^{\pm 2i\pi/4}$, are associated to the set of parameters (\ref{eq:Picard}). We already considered this case at the end of Section \ref{sec:st_transformation}. By conjugating all the matrices by $M_t$, one has
$(\sigma_{0t},\sigma_{1t})\mapsto(-\sigma_{0t},-\sigma_{1t})$, which is the same as \eqref{eq:identification}.  

It is convenient to represent a point $P_{Picard}\in \mathcal{M}_{0,4}^{SU(2)}(\bs \theta_{Picard})$ by using the variable $\sigma_{1t}$ instead of $s_{0t}$,  
$$P_{Picard}=(\bs \theta_{Picard},\sigma_{0t},\sigma_{1t}).$$ 
The relation between $s_{0t}$ and $\sigma_{1t}$ greatly simplifies for $\mathcal{M}_{0,4}^{SU(2)}(\bs \theta_{Picard})$. From the expression of  $p_{1t}(=2\cos 2\pi \sigma_{1t})$ as given in   (\ref{eq:p0t}) and from  \eqref{eq:involution}, one obtains:
\eq{
\label{eq:sots1t}
 \text{for}\;\; \bs \theta= \bs \theta_{Picard}: \quad s_{0t}=-e^{2\pi i \sigma_{1t}}, \quad \iota'(s_{0t})=-e^{-2\pi i\sigma_{1t}} 
}
Accordingly, the closed 2--form (\ref{eq:Darboux}) in the new coordinates simply reads:
\eq{
\frac{1}{2\pi i} d\sigma_{0t} \wedge \frac{d s_{0t}}{s_{0t}}=d\sigma_{0t} \wedge d \sigma_{1t}
}

We focus first on the  function  $\taut(P_{Picard},t,\bar t)$, see the (\ref{eq:taut}). The associated  structure constants take the simple form:
\eq{
C(\bs\theta_{Picard};\sigma_{0t})=\frac{c^+}{\cos(\pi\sigma_{0t})}16^{-\sigma_{0t}^2},\quad 
C(-\bs\theta_{Picard};\sigma_{0t})=c^- \cos(\pi\sigma_{0t}) 16^{-\sigma_{0t}^2}.
}
Using the above result, one can express  $\taut(P_{Picard},t,\bar t)$  via the following double sum:
\eq{
\taut(P_{Picard},t,\bar{t})=\frac{c^+ c^-}{|t^{\frac18}(1-t)^{\frac18}\vartheta_3(0|\eta(t))|^2}
\Sum_{(n,m)\in \mathbb{Z}^2}e^{2\pi i\sigma_{1t}(n-m)} q^{(\sigma_{0t}+n)^2}\bar q^{(\sigma_{0t}+m)^2},
}
where the (\ref{eq:Picardblock}) and (\ref{eq:sots1t}) have been used. Henceforth, we will neglect $c^+c^-$ factor.

We can study now the weakly invariant measures on $\mathcal{M}_{0,4}^{SU(2)}(\bs \theta_{Picard})$. The action  (\ref{eq:braidaction}) of the braid group on $\mathcal{M}_{0,4}^{SU(2)}(\bs \theta_{Picard})$ reads
\eq{
b_{0t}(M_0)\sim M_0,\quad b_{0t}(M_t)\sim M_t,\quad
b_{0t}(M_1)\sim \bpm 0&ie^{-2\pi i(\sigma_{1t}-\sigma_{0t})}\\ie^{2\pi i(\sigma_{1t}-\sigma_{0t})}&0\epm\,.
}
Notice that to obtain the above equations we applied  an appropriate simultaneous conjugation. A point $P_{Picard}$, under the action of $b_{0t}$ and $b_{t1}$ transforms in the following way:
\eq{
b_{0t}(P_{Picard})=(\bs \theta_{Picard},\sigma_{0t},\sigma_{1t}-\sigma_{0t})\,,\\
b_{t1}(P_{Picard})=(\bs \theta_{Picard},\sigma_{0t}+\sigma_{1t},\sigma_{1t}).
}
These two transformations generate the $SL(2,\mathbb Z)$ group
\eq{
\bpm \sigma_{0t}\\\sigma_{1t}\epm\mapsto \bpm a&b\\c&d\epm\bpm\sigma_{0t}\\\sigma_{1t}\epm\,,\qquad
\begin{vmatrix}a&b\\c&d\end{vmatrix}=1\,.
\label{eq:SL2}
}

We know two possibilities to construct distribution, invariant under these transformations. 

\subsubsection{Continuous spectrum: back to the Runkel-Watts theory} 
The first one is very universal: we just take
$d\mu^{unitary}_{\bs\theta_{Picard}}$ given by \eqref{eq:dmu_unitary}:
\eq{
d\mu^{unitary}_{\bs\theta_{Picard}}=d\sigma_{0t}\wedge d\sigma_{1t}\cdot \delta({\rm Im\,}\sigma_{0t})\delta({\rm Im\,}\sigma_{1t})
}
Using this distribution  we get
\eq{
\mc F(t,\bar t)[d\mu^{unitary}_{\bs\theta_{Picard}}]=\int d\sigma_{0t}\wedge d\sigma_{1t}\; \taut(P_{Picard};t,\bar t)=\\=
\Sum_{n\in\mathbb Z}\int_{0}^1 \; d\sigma_{0t}\;\frac{ e^{-2\pi(\sigma_{0t}+n)^2 {\rm Im}\; \eta(t)}}{|t^{\frac18}(1-t)^{\frac18}\vartheta_3(0|\eta(t))|^2}=
\int_{-\infty}^\infty\; d\sigma \frac{ e^{-2\pi \sigma^2 {\rm Im} \;\eta(t)}}{|t^{\frac18}(1-t)^{\frac18}\vartheta_3(0|\eta(t))|^2}=\\=
|t^{\frac18}(1-t)^{\frac18}\vartheta_3(0|\eta(t))|^{-2}(2 {\rm Im}\; \eta(t))^{-\frac12}
}
The above correlation function, that can be found in \cite{zz90}, is a   Runkel-Watts correlator, see Eq.45 in \cite{rw01}.

\subsubsection{Discrete spectrum: Ashkin-Teller spin correlators} 

Let us consider some vector $(\omega_1,\omega_2)$  and its $SL(2,\mathbb Z)$ orbit, which we denote
\eq{
O_{(\omega_1,\omega_2)}=SL(2,\mathbb Z)\cdot (\omega_1,\omega_2)\,.
}
The second possibility to build an invariant measure is to define it as  supported by $O_{(\omega_1,\omega_2)}$:
\eq{
d\mu^{O_{(\omega_1,\omega_2)}}=\Sum_{\vec x\in O_{(\omega_1,\omega_2)}} \delta(\sigma_{0t}-x_1)\delta(\sigma_{1t}-x_2)
}
We have the following
\begin{proposition}
All finite orbits $O_{(\omega_1,\omega_2)}$ coincide with some $O_{(0,1/p)}=O_{1/p}$ for $p\in\mathbb Z$, which, in its turn, can be obtained by
additions and set-theoretic subtractions of the finite lattices 
\eq{
L_{1/p}=\frac1{p}(\mathbb Z/p\mathbb Z)^2\subset (\mathbb Q/\mathbb Z)^2\subset (\mathbb R/\mathbb Z)^2\,,
}
so the basis of invariant measures is given by measures, supported by $L_{1/p}$.
\end{proposition}

Proof of this proposition goes in the following way. First we check that any vector $(\omega_1,\omega_2)$ is equivalent to $(\omega,0)$. Actually,
if there is $(\omega_1,\omega_2)$ with $|\omega_1|\leq |\omega_2|$, then we may transform it to $(\omega_1,\omega_2\pm\omega_1)$ --- this procedure
necessarily decreases $\min(|\omega_1|,|\omega_2|)$, and finally we come to $(\omega,0)$~\footnote{described procedure is nothing but the Euclid's algorithm}. Then $\omega$ should be rational number $\omega=p'/p$~\footnote{$p$ and $p'$ are not necessarily prime}, otherwise
we get infinite orbit.

Obviously we have an inclusion $O_{p'/p}\subset L_{1/p}$. Now we wish to describe the decomposition of $L_{1/p}$ into the union of orbits.
This is the same as to describe the orbits of $SL(2,\mathbb Z)\ltimes (p\mathbb Z)^2$ acting on $\mathbb Z^2$: $(A,x,y):\frac1p(a,c)\mapsto A\cdot\frac1p(a+p x,c+p y)$.
It is clear that the number $d=\gcd(a,c,p)$ is invariant under such action.
Now we show that it is the only invariant, so the point $\frac1p(a,c)$ belongs the orbit $O_{\frac1p \gcd(a,c,p)}$. Moreover, we show that $O_{\tilde d'/\tilde p}=O_{1/\tilde p}$ if $\tilde d'/\tilde p$ is irreducible, so
\eq{
\label{eq:orbit_decomposition}
L_{1/p}=\bigsqcup_{\tilde p|p}O_{1/\tilde p}\,.
}

Actually, first we can map point $\frac1p(a,c)$ to $\frac1p\left(0,d'\right)$ using $SL(2,\mathbb Z)$ action to perform the Euclid's algorithm. Here $d'=\gcd(a,c)$.
Then we reduce fraction $d'/p=\tilde d'/\tilde p$. Since $\gcd(\tilde d',\tilde p)=1$, one can find such $\alpha, \beta\in\mathbb Z$ that $\alpha\tilde d'-\beta\tilde p=1$ and then construct  $SL(2,\mathbb Z)$ matrix $A$:
\eq{
A=\bpm\tilde d'&\tilde p\\\beta&\alpha\epm\,,\quad A\bpm 0\\\tilde d'\epm = \bpm \tilde p\tilde d'\\1+\beta\tilde p\epm 
}
In this way we have shown that any orbit in $L_{1/p}$ is equivalent to $O_{1/\tilde p}$, where $\tilde p$ is some divisor of $p$, so we proved \eqref{eq:orbit_decomposition}.

Now we introduce the invariant distribution
\eq{
d\mu^{1/p}=\Sum_{a,c=0}^{p-1}\delta(\sigma_{0t}-a/p)\delta(\sigma_{0t}-c/p)\,.
}
Using \eqref{eq:orbit_decomposition} we write
\eq{
d\mu^{1/p}=\Sum_{\tilde p|p}d\mu^{O_{1/\tilde p}}\,,
}
and then using M\"obius inversion formula we get
\eq{
d\mu^{O_{1/p}}=\Sum_{\tilde p|p}\mu\left(\frac{\tilde p}{p}\right)d\mu^{1/\tilde p}\,,
}
where $\mu\left(\frac{\tilde p}{p}\right)$ is the M\"obius function. Therefore any invariant measure can be expressed in terms of $d\mu^{1/\tilde p}$. \hfill $\square$

Now we start to study such measures, $d\mu^{1/p}$, and first we study $d\mu^0$, which gives 
\eq{
\mc F(t,\bar t)[d\mu^0]=\taut(\bs\theta_{Picard};0,0;t,\bar t)=|t|^{-\frac14}|1-t|^{-\frac14}\,.
}
So it is just the free field correlator.

Moving to the general case we make two updates: first, for the technical reasons we can consider only even $p=2N$, and second, we can consider three more measures which are not $SL(2,\mathbb Z)$ invariant, but instead form its 3-element orbit:
\eq{
d\mu^{\frac1{2N},\epsilon,\epsilon'}=\Sum_{a,c=0}^{2N-1}\delta\left(\sigma_{0t}-\frac{a+\epsilon/2}{2N}\right)\delta\left(\sigma_{1t}-\frac{c+\epsilon'/2}{2N}\right)\,,
\label{eq:measure_at}
}
where $\epsilon,\epsilon'\in \mathbb Z/2\mathbb Z$. We see that the measure $d\mu^{\frac1{2N},0,0}$ is invariant, but three other measures $d\mu^{\frac1{2N},1,0}=:d\mu^{\frac1{2N},1}$, $d\mu^{\frac1{2N},0,1}=:d\mu^{\frac1{2N},2}$ and $d\mu^{\frac1{2N},1,1}=:d\mu^{\frac1{2N},3}$
are permuted by $SL(2,\mathbb Z/2\mathbb Z)=S_3$. The two $SL(2,\mathbb Z)$ generators $S=\bpm 0&1\\-1&0\epm$ and $T=\bpm 1&1\\0&1\epm$ map to permutations as follows:
\eq{
S\mapsto (12)\,,\quad T\mapsto (23)\,.
}
To get correlation function first one has to compute the sum
\eq{
\Sum_{m,n\in\mathbb Z}\Sum_{a,c=0}^{2N-1}e^{\pi i\frac{c+\epsilon'/2}{N}(n-m)} q^{(\frac{a+\epsilon/2}{p}+n)^2}\bar q^{(\frac{a+\epsilon/2}{2N}+m)^2}=
\\
=
\Sum_{n,k\in\mathbb Z}(-1)^{\epsilon'k}\Sum_{a=0}^{2N-1}q^{(\frac{a+\epsilon/2}{2N}+n)^2}\bar q^{(\frac{a+\epsilon/2}{2N}+n+2Nk)^2}=\\=
\Sum_{n,k\in\mathbb Z}(-1)^{\epsilon'k}q^{(\frac{n}{2N})^2}\bar q^{(\frac{n}{2N}+2Nk)^2}=
\Sum_{n,k\in\mathbb Z}(-1)^{\epsilon'k}q^{(\frac{n+\epsilon/2}{2N}+Nk)^2}\bar q^{(\frac{n+\epsilon/2}{2N}-Nk)^2}/,.
}
 Using this sum the correlation function is rewritten as
 \eq{
 \mc F(t,\bar t)[d\mu^{\frac1{2N},\epsilon,\epsilon'}]=\frac{|t|^{-\frac14}|(1-t)|^{-\frac14}}{\vartheta_3(0|\eta(t))|^{2}}
\Sum_{n,k\in\mathbb Z}(-1)^{\epsilon'k}q^{(Nk+\frac{n+\epsilon/2}{2N})^2}\bar q^{(Nk-\frac{n+\epsilon/2}{2N})^2}
}
that corresponds to $\mathcal{F}_{\bs \theta_{Picard}}(t,\bar{t})$ defined previously in (\ref{eq:ATcorr}).

\subsection{Analytic Liouville theory}
\label{sec:Analytic}

In the previous examples the weakly invariant distributions were actually just invariant.
Here we consider the situation where the second example of weakly invariant distribution is realized: namely, in this case we take an integral over some 2-cycle, such that its homology class is invariant under the braiding transformations.
It turns out that so constructed distribution provides the correlation functions of the analytic Liouville theory \cite{RiSa15}.
In this way we prove crossing invariance of this theory, conjectured in \cite{RiSa15}.

Consider the distribution concentrated on the contour $\mc C_{h,r}$ \eqref{eq:Chr}:
\eq{
d\mu^{analytic(h,r)}_{\bs\theta}=\frac1{2\pi i}d\sigma_{0t}\wedge\frac{ds_{0t}}{s_{0t}}\cdot
\delta(\operatorname{Im} \sigma_{0t}-h)\delta(|s_{0t}|-e^{-2\pi r})\,,
\label{eq:dmu_analityc}}
where $h\neq0$ and the associated function:
\eq{
\mc F(t,\bar{t})[d\mu^{analytic(h,r)}_{\bs\theta}]=\int_{ih}^{ih+1}
d\sigma_{0t} \Int_{|s_{0t}|=e^{-2\pi r}}
\frac{ds_{0t}}{2\pi i s_{0t}}\;\taut(P,t,\bar t)\
}
Using the fact that nothing depends on $|s_{0t}|$, the integration over $s_{0t}$ follows strictly the one in \eqref{eq:int_s}. We obtain:
 \eq{
\mc  F(t,\bar{t})[d\mu^{analytic(h,r)}_{\bs\theta}]=\Sum_{n\in\mathbb Z}\int_{ih}^{ih+1} d\sigma_{0t}
\Phi(\theta_0,\theta_t,\sigma_{0t}+n)\Phi(\theta_\infty,\theta_1,\sigma_{0t}+n)
\times\\\times \mc B(\bs\theta,\sigma_{0t}+n,t) \mc B(\bs\theta,\sigma_{0t}+n,\bar t)\\
=\int_{ih+\mathbb R} d\sigma_{0t}
\Phi(\theta_0,\theta_t,\sigma_{0t})\Phi(\theta_\infty,\theta_1,\sigma_{0t})\mc B(\bs\theta,\sigma_{0t},t) \mc B(\bs\theta,\sigma_{0t},\bar t),
\label{eq:int_analytic}
}
that, comparing with (\ref{eq:AL}) shows that:
$$\mc F(t,\bar{t})[d\mu^{analytic(h,r)}_{\bs\theta}]= \mathcal{F}^{AL}_{\bs \theta}(t,\bar{t}).$$

Note that the above formulas are very similar to the ones we saw in the case of $d\mu^{unitary}_{\bs\theta}$ \eqref{eq:dmu_unitary}, since in the both cases we integrate the form $d\omega$ over submanifold. However, as we already said, there is an important difference: in the unitary case the submanifold is invariant under the braid group action, while in the analytic case this is not true. Let's see this more in detail.

We want to prove crossing invariance and single-valuedness of $\mathcal{F}^{AL}_{\bs \theta}(t,\bar{t})$ by observing that the homology class of the submanifold of integration in (\ref{eq:dmu_analityc}) is preserved by the braid group. We use the following strategy:
\begin{itemize}
\item $A$: we check that the singularities of the integrand do not affect invariance of the integral.
  Namely, even when we move contour (\ref{eq:int_analytic}) through the pole of the tau function, integral does not change because corresponding residue vanishes.
  In other words, one does not have to add extra cycles encircling singularities in the finite domain.
  So the only thing that we have to check is that integration cycle lies in the same homology class in $H_2((\mathbb C^\times)^2,\mathbb Z)=\mathbb Z$ after the transformation, where by $(\mathbb C^\times)^2$ we denote a manifold with coordinates $(e^{2\pi i\sigma_{0t}},s_{0t})$.
  
\item $B$: we check that homology class of the integration contour $[\mc C_{h,r}]$ is invariant.
\end{itemize}

Let's consider the point $A$. The integrand is analytic in $\sigma_{0t}$ and $s_{0t}$ everywhere except for $\sigma_{0t}\in\{0,\frac12\}$ (or, considering the  
the real line, everywhere except for $\sigma_{0t}\in\frac12\mathbb Z$).
We have to verify that moving the contour through these points does not change the value of the integral: therefore,  even if contour goes through the singular point, it can be correctly defined by shifting in one of the two directions, i.e. with $h<0$ or $h>0$.
One observes that the two residues, at $\sigma_{0t}=0$ and  at $\sigma_{0t}=1/2$,  vanish due to the following relations:
\eq{
\Sum_{n\in\mathbb Z}\operatorname{Res}_{\sigma_{0t}=n} \Phi(\theta_0,\theta_t,\sigma_{0t})\Phi(\theta_\infty,\theta_1,\sigma_{0t})
\mc B(\bs\theta,\sigma_{0t},t) \mc B(\bs\theta,\sigma_{0t},\bar t)d\sigma_{0t}=0\,,\\
\Sum_{n\in\mathbb Z}\operatorname{Res}_{\sigma_{0t}=n+\frac12} \Phi(\theta_0,\theta_t,\sigma_{0t})\Phi(\theta_\infty,\theta_1,\sigma_{0t})
\mc B(\bs\theta,\sigma_{0t},t) \mc B(\bs\theta,\sigma_{0t},\bar t)d\sigma_{0t}=0\,.
\label{eq:zero_residue}
}
To prove this we notice first that all the functions under the sum are even functions of $\sigma_{0t}$. This property is manifest in the definition of the constants (\ref{eq:Phidef}) and, concerning the conformal blocks $\mc B(\bs\theta,\sigma_{0t},t)$ and $\mc B(\bs\theta,\sigma_{0t},\bar{t})$, we recall that they depend only on the dimension $\sigma_{0t}^2$ of the intermediate channel. Now suppose that there is some even function
$f(\sigma)=f(-\sigma)$. Then the two functions
\eq{
\tilde f_0(\sigma)=\Sum_{n\in\mathbb Z}f(\sigma+n)\qquad\text{and}\qquad \tilde f_{\frac12}(\sigma)=\Sum_{n\in\mathbb Z}f(\sigma+n+\frac12)
}
are $\mathbb Z$-periodic and even: $\hat f_0(\sigma)=\hat f_0(-\sigma)$, $\hat f_{\frac12}(\sigma)=\hat f_{\frac12}(-\sigma)$. This implies automatically that
$\operatorname{Res}_{\sigma=0}\hat f_0(\sigma)d\sigma=\operatorname{Res}_{\sigma=0}\hat f_{\frac12}(\sigma)d\sigma=0$, and proves thus \eqref{eq:zero_residue}.
In this way we actually proved the point $A$ above: singularities in the finite domain do not affect invariance of the integral. We also proved actually that the analytic distribution does not depend on $h$ and $r$ in a sense of weak equivalence:
\eq{
d\mu^{analytic(h,r)}_{\bs\theta}\simeq d\mu^{analytic(h',r')}_{\bs\theta}\simeq d\mu^{analytic}_{\bs\theta}\,.
\label{eq:analytic_equivalence}
}
One interprets this fact as effectively we do not have holes in the finite domain, so $[\mc C_{h,r}]=[\mc C_{-h,r}]=[\mc C_{h,-r}]=[\mc C_{-h,-r}]$, and homology class does not depend on $(h,r)$:
\eq{
  \left[\mc C_{h,r}\right]=\left[\mc C_{h',r'}\right]
\label{eq:hr_independence}
}

We can now pass to point $B$ of the proof. We have already checked in Section \ref{sec:st_examples} that in the limit $|h|\to\infty, |r|\to\infty, |h-r|\to\infty$ homology class $[\mc C_{h,r}]$ transforms to $[\mc C_{\tilde h,\tilde r}]$.
Because of \eqref{eq:hr_independence}, first, it implies invariance of $[\mc C_{h,r}]$ in the limit, and second, by the deformation argument~\footnote{One can first pull initial contour to infinity, apply crossing transformation, and then pull it back. Notice that in the previous version of the paper, we computed the effects on the transformations in a region infinitesimal close to the Picard point $\bs\theta_{\text{Picard}}$ and then we used an analytic continuation argument to extend the result to all values of $\bs\theta$. As the referee pointed out, this proof actually was not completed.} it implies invariance for all $(h,r)$. 

There is also simpler braiding transformation \eqref{eq:ttbraiding} acting by $b_{0t}^2: s_{0t}\mapsto s_{0t} e^{4\pi i \sigma_{0t}}$.
It maps $\mc C_{h,r}$ to $\mc C_{h,r+2h}$, so also preserves $[\mc C_{h,r}]$.
The same also holds for half-braiding $b_{0t}$ \eqref{eq:half_brading}. So all possible transformations preserve $[\mc C_{h,r}]$, and this completes the proof.

In this way we have proved  that correlation function in analytic Liouville theory \eqref{eq:int_analytic} is $h$-independent,
crossing-invariant and single-valued.

\section{Conclusions}
In this paper we discussed the construction of crossing invariant correlation functions in $c=1$ conformal field theories. Using the relation between isomonodromic deformations and Virasoro conformal blocks, we proposed the tau functions instead of the conformal blocks as a basis to construct crossing invariant correlation functions. The basic idea is that the transformation properties of the tau function under the braid group are simpler than the ones of the conformal blocks, in particular when a continuous spectrum is considered. We proposed the ansatz (\ref{eq:F}) for crossing invariant correlation function according to which the problem of constructing of a class of CFT solutions reduces to the definition of weakly invariant distributions on the moduli space $\monSpace$. We presented four examples of such distributions, given in (\ref{eq:distr_gff}), (\ref{eq:dmu_unitary}), (\ref{eq:measure_at}) and (\ref{eq:dmu_analityc}). We showed that these four distributions provide respectively the correlation functions of the free Gaussian, of the Runkel-Watts, of the Ashkin-Teller and of the analytic Liouville theory.  

We showed that our approach permits not only to recover very different theories by using a common framework. More important, we could prove crossing symmetry for the analytic Liouville theory, a case where other methods, based for instance on the Teschner-Ponsot formulas \cite{pt99} seem too complicated for this value of the central charge. Nevertheless, there are crucial issues that need to be investigated further.
We expect indeed that not all the possible $c=1$ CFT solutions are in the form (\ref{eq:F}); for instance, this form seems not designed to describe correlations with non-diagonal external fields. And, even assuming the form (\ref{eq:F}), we cannot characterize all its solutions as we do not know how to classify completely the weakly invariant distributions on $\monSpace$. 

There is also a series of questions and problems for future investigations:

\begin{enumerate}

\item One obvious way to construct invariant measures on $\monSpace$ is to take sums of delta-functions over the finite orbits of the action of the braid group: $d\mu(P)=\Sum_{P'\in O}\delta(P,P')$.
All such orbits were classified in \cite{LTy08}, corresponding solutions of Painlev\'e  VI are also known.
Most of these finite orbits arise from monodromies corresponding to finite subgroups $G\subset SU(2)$ --- 
conjecturally, these cases should correspond to $c=1$ CFT's on orbifolds, and we should be able to construct closed formulas for correlation functions in such theories.
More interesting question is what is the CFT meaning of the orbits that do not come from the finite subgroups.

\item Despite the fact that the Ashkin-Teller correlation functions are known, this case needs better understanding from the methodological point of view.
At the moment we are able to reproduce only the theories with integer parameter $N$ by taking average over the finite orbit:
the question is how this procedure has to be modified in the case of arbitrary $N$.

\item As in the Virasoro case, recent efforts have been focused in finding new boostrap solutions in Toda theories \cite{DEI18}. Definitely, our construction should work for the $N\times N$ case related to $W_N$ algebras with $c=N-1$, since relation between higher-rank isomonodromic deformations and W-algebras is also known \cite{Gav15}. 
We conjecture that in this case there should be the analog of \eqref{eq:chi_product}, stating that the product of two connection constants equals to the ratio of 3-point functions.
As a consequence, one should get four types of $c=N-1$ theories with $W_N$ symmetry: 1) analog of the Runkel-Watts theory --- the limit of $W_N$ unitary minimal models with the fusion rules governed by existence of solutions of equation $M_1M_3M_3=1$ with fixed conjugacy classes of $M_\nu$; 2) analog of the analytic Liouville theory; 3) analog of the Ashkin-Teller model; 4) analogs of the orbifold CFT's.
Theory of the first type already appeared, for instance, in \cite{fre10}.
Some steps towards the study of the orbits of the braid group were already done in \cite{GIL18}, where the action of three involutions on the monodromy data is written.

\item An approach worth to explore is the construction of  weakly invariant distributions from the study of the (linear) action of the braid group on cohomologies of $\monSpace$ along the lines of \cite{IwasakiUehara06} and \cite{CantatLoray07}.

\item  The singularities of the tau functions at $\sigma_{0t}\in\frac12\mathbb Z$ should be considered more in detail. There might be some construction in which the integral is ``concentrated'' around these submanifolds: we expect this mechanism to produce some correlation functions in the logarithmic $c=1$ CFT.

\item There are several fundamental questions that arise concerning the general development of our approach. At the moment we are able to give some functions that are crossing invariant almost by construction. To guarantee that such functions are actual correlation functions in CFT, one should be able to represent them in a form of bilinear combinations of conformal blocks with factors \emph{decomposed into products of 3-point functions}, like \eqref{eq:ATcorr}, \eqref{eq:RW_correlator}, \eqref{eq:AL}.
As a matter of fact, the functions we constructed in this paper turned out to fulfill this requirement.  However it would be very interesting to reformulate this factorization property as some additional property on $d\mu_{\bs\theta}(P)$.

There is also the question whether the ansatz \eqref{eq:F} is the most general expression that gives crossing invariant functions, or there can be some more complicated ``non-local'' integrals. 
Finally, we stress that, in our approach, only diagonal fields are allowed in the external channels (but arbitrary fields in the internal channel).
So the question is how one can modify this construction, and which properties of the connection constant should be used to allow  general fields in all channels.

If we wish to reformulate the problem of classification of $c=1$ crossing invariant CFTs as some problem about weakly invariant distributions, all these questions have to be answered.

\end{enumerate}

We hope to return to these problems elsewhere.

\section{Acknowledgements}

We are grateful to M.~Bershtein, B.~ Estienne, N.~Iorgov, O.~Lisovyy, A.~Litvinov, A.~Marshakov for their interest to this work and for discussions, and especially to S.~Ribault, for careful reading of the text and for many useful comments and questions. We thank Sylvain Ribault for pointing out the issue of the factorization property of our ansatz. We also thank the referee for pointing out issues in the first version of the proof in Section \ref{sec:Analytic}, see footnote in this Section. We are also grateful to the organizers of the conference ``Random Geometry and Physics'', which was held in Paris in 2016, where this work was initiated. We are grateful to the University of Tours, and especially to O.~Lisovyy, for his hospitality during the workshop in 2017 where the first results of this work were reported.

The research of PG was carried out within the HSE University Basic Research Program and funded jointly by the Russian Academic Excellence Project '5-100', by RFBR grant mol\_a\_ved 18-31-20062, and by Young Russian Mathematics award.

\bibliographystyle{JHEP}
\bibliography{bibpainleve.bib}

\end{document}